\documentclass[useAMS,usenatbib]{mn2e}
\usepackage{color}
\usepackage{graphicx}
\usepackage{lscape}
\usepackage{amssymb,amsmath}
\usepackage{pifont}
\usepackage{url}
\usepackage{float}

\title[Non-linearity in the P-L, P-W and P-C relations]
{Large Magellanic Cloud Near-Infrared Synoptic Survey. III. 
A Statistical Study of Non-Linearity in the Leavitt Laws}
\author[Bhardwaj et al.]{Anupam Bhardwaj$^1$\thanks{E-mail:
anupam.bhardwajj@gmail.com}, Shashi M. Kanbur$^2$, Lucas M. Macri$^3$, Harinder P. Singh$^1$, 
\newauthor 
Chow-Choong Ngeow$^4$ and Emille E. O. Ishida$^5$\\
\vspace{1pt}\\
1. Department of Physics \& Astrophysics, University of Delhi, Delhi 110007, India. \\
2. State University of New York, Oswego, NY 13126, USA.\\
3. Mitchell Institute for Fundamental Physics \& Astronomy, Department of Physics \& Astronomy, Texas A\&M University, \\
~~~College Station, TX 77843, USA \\
4. Graduate Institute of Astronomy, National Central University, Jhongli 32001, Taiwan \\
5. Max-Planck-Institut f{\"u}r Astrophysik, Karl-Schwarzschild-Stra\ss e 1, 85748, Garching, Germany\\
}
\begin{document}

\date{Accepted 2016 January 5. Received 2015 December 27; in original form 2015 November 2}

\pagerange{\pageref{firstpage}--\pageref{lastpage}} \pubyear{2015}

\maketitle

\label{firstpage}

\begin{abstract}
We present a detailed statistical analysis of possible non-linearities in the Period-Luminosity (P-L), Period-Wesenheit (P-W) and Period-Color (P-C) relations for Cepheid variables in the LMC at optical ($VI$) and near-infrared ($JHK_{s}$)  wavelengths. We test for the presence of possible non-linearities and determine their statistical significance by applying a variety of robust statistical tests ($F$-test, Random-Walk, Testimator and the Davies test) to optical data from OGLE III and  near-infrared data from LMCNISS. For fundamental-mode Cepheids, we find that the optical P-L, P-W and P-C relations are non-linear at 10 days. The near-infrared P-L and the $W^H_{V,I}$ relations are non-linear around 18 days; this  break is attributed to a distinct variation in mean Fourier amplitude parameters near this period for longer wavelengths  as compared to optical bands. The near-infrared P-W relations are also non-linear except for the $W_{H,K_s}$ relation. For first-overtone mode Cepheids, a significant change in the slope of P-L, P-W and P-C relations is found around 2.5 days only at optical wavelengths. We determine a global slope of $\textrm{-}3.212\pm0.013$ for the $W^H_{V,I}$ relation by combining our LMC data with observations of Cepheids in Supernovae host galaxies \citep{riess11}. We find this slope to be consistent with the corresponding LMC relation at short periods, and significantly different to the long-period value. We do not find any significant difference in the slope of the global-fit solution using a linear or non-linear LMC P-L relation as calibrator, but the linear version provides a $2\times$ better constraint on the slope and metallicity coefficient.

\end{abstract}


\section{Introduction}
\label{sec:intro}

Cepheids are pulsating variables with a well-defined absolute magnitude for a given 
period and hence exhibit a Period-Luminosity (P-L) relation also known as the Leavitt 
Law \citep{leavitt}. Thus, Cepheids are used as the standard candles and are essential 
component in determining extra-galactic distances. The non-linearity of the fundamental-mode 
Cepheid Period-Luminosity relation has been a subject of many studies in the past decade 
\citep{tammann03, tammann04, smkc05, ccn05, choong06, smkccn05, tammann09, varela13}. 
Theoretical models were also developed to analyse Period-Luminosity, Period-Color (P-C) 
and Period-Luminosity-Color (PLC) relations \citep{bono99, smk2, smk3, smk5}.
A comparison of the slopes of P-L and P-C relations in the Galaxy and the Large \& Small 
Magellanic Clouds (LMC, SMC) was provided by \citet{tammann03}. Similarly, studies of the 
P-L and P-C relations in the LMC were carried out first by \citet{tammann04}
and investigated further with a rigorous statistical approach in a series of papers 
\citep{smkc05, ccn05, smkccn05, choong06, shashi07}. These authors also extended the 
analysis to study breaks in P-L, P-C and Amplitude-Color (A-C) relations 
at various phases of pulsation \citep[][and the references therein]{smk1, smk3, smk6, bhardwaj14}.
A study at $BVI_{c}JHK_{s}$ wavelengths was carried out by \citet{cchoong08}, to test the non-linearity 
in the LMC P-L relations. The counterparts of OGLE-III Cepheids in 2MASS and SAGE catalogs 
were used to derive P-L relations, and to test for possible non-linearity at 10 days in \citet{ngeow09}.

In most previous studies \citep{tammann03, tammann04, tammann09}, the test for 
non-linearities consisted of fitting slopes to data on opposite sides of a break 
point, deriving their uncertainties, and evaluating their difference in terms of 
standard deviations. \citet{shashi07} have shown rigorously that this procedure 
necessarily leads to a greater probability of error than using a statistical test 
such as the $F$-test. Thus, this paper concentrates on applying rigorous statistical 
tests to mean light relations and compares these results to previous work that has 
studied these non-linearities at different phases. Many researchers have advocated 
the use of longer wavelengths in P-L distance work since these relations have a lower 
dependence on metallicity and extinction, a smaller intrinsic dispersion and are supposedly linear 
\citep{monson12, ripepi12, inno13}. For these reasons, we extend our rigorous tests to 
near-infrared wavelengths.

\citet{smk2, smk3, smk5} studied theoretical nonlinear models of Cepheids with a view 
of understanding the interaction between the photosphere and hydrogen ionization front 
and how this may cause possible sharp non-linearities in the P-C relation at various phases. 
They found that there are distinct phases and period ranges for which this interaction
produces a sharp change to the P-C relation. Recently, \citet{bhardwaj14} have discussed 
various non-linearities in the P-C and A-C relations for Cepheids in the Milky Way and 
Magellanic Clouds at the phases of maximum and minimum light. These results were found to be 
consistent with the theoretical explanations \citep{smk93, smk2, smk3, smk5} relating
the hydrogen ionization front and the stellar photosphere and the properties of the Saha ionization
equation. The P-L and P-C relations at mean light are usually obtained by the numerical average of 
the corresponding magnitudes and colors at every phase during a pulsation period. 
Therefore, any changes in P-C and P-L relations at particular phases during the pulsation cycle 
can indeed have effects on the mean light P-C and P-L relation. P-C relations can affect the 
P-L relations through PLC relations and hence it is essential to investigate the various 
non-linearities in P-L and P-C relations at mean light. We also include reddening independent 
P-W relations in our analysis.

Motivated by the aforementioned reasons, we use light curve data having full phase coverage 
to extend our work for the first time to investigate observational evidence for non-linearity 
in near-infrared P-L, P-W and P-C relations. We combine optical data from the OGLE-III survey 
\citep{oglelmcceph, ulac13} with data from the Large Magellanic Cloud Near-Infrared Synoptic 
Survey \citep[LMCNISS,][{ and erratum}, hereafter, Paper I]{macri14}. We use the $F$-test, random walk, testimator and Davies  
tests to determine possibly statistically significant breaks in the P-L, P-W and P-C relations 
for fundamental and first-overtone mode Cepheids. This paper presented as the third in the series, 
extends the work of Paper I and \citet[][hereafter, Paper II]{aas15} and uses 
the catalog to study non-linearity in the Leavitt Law. The near-infrared P-L and P-W relations
derived in Paper I and II based on our data are very important in distance scale applications
and, therefore, a rigorous analysis of non-linearity is essential to avoid any discrepancies in Cepheid
based distance estimates.

The paper is structured as follows. In Section~\ref{sec:lc_data}, we present details of the photometric 
data and extinction corrections. In Section~\ref{sec:method}, we describe the various statistical methods 
used to test the presence and significance of non-linearities in the P-L, P-W and P-C relations. In 
Section~\ref{sec:analysis}, we present the results of our analysis. We also discuss the impact of observed 
non-linearities on the distance scale. We summarize all the results and the conclusions
from this study in Section~\ref{sec:discuss}.


\section{Data and Extinction Correction}
\label{sec:lc_data}

The photometric mean magnitudes for optical $V$- and $I$-band for Cepheids in the LMC are taken 
from the Optical Gravitational Lensing Experiment (OGLE - III) survey \citep{oglelmcceph}.  
There are 1849 fundamental (FU) and 1238 first-overtone (FO) mode Cepheids in the LMC. The optical 
bands sample will be referred to as ``OGLE-III'' in this work. We augment this sample with mean magnitudes 
for 26 long period Cepheids in the LMC from the OGLE-III Shallow Survey \citep{ulac13}. The photometric 
system is exactly similar for both datasets and hence the combined sample will be referred to as ``OGLE-III-SS''.

The mean magnitudes for the LMC in the optical bands are corrected for reddening using the maps of 
\citet{hasch11}. The color excess $E(V\textrm{-}I)$ values are obtained from these Haschke maps using RA/DEC 
for Cepheids in the LMC taken from the OGLE-III database\footnote{\url{http://dc.zah.uni-heidelberg.de/mcextinct/q/cone/info}}. 
We corrected mean magnitudes using the \citet{card89} extinction law $A_{\lambda} = R_{\lambda}E(B\textrm{-}V)$, 
where $E(V\textrm{-}I)$ is related to $E(B\textrm{-}V)$ by the relation $E(V\textrm{-}I) = 1.38 E(B\textrm{-}V)$ 
\citep{tammann03}. For the LMC, the values for total-to-selective absorption ($R_{\lambda}$) are $R_{V} = 2.40, R_{I} = 1.41$,
corresponding to $E(V\textrm{-}I)$ color excess values.

We use our recently released data from the LMC Near-infrared Synoptic Survey to determine the 
significance of possible non-linearities in these relations for FU and FO mode Cepheids. 
The observations for these Cepheids along with their calibration into the 2MASS photometric system, 
adopted reddening law and extinction corrections are discussed in detail in Paper I.
The final sample used in calibrating P-L relations includes 775 FU and 474 FO mode Cepheids in the LMC, which was
used to derive P-W relations in Paper II.

\section{The Statistical Tests}
\label{sec:method}

\subsection{$F$-test}
Our interest for the present study is to test whether the P-L, P-W and P-C relations display statistically significant 
evidence of a change of slope. We make use of the method adopted by \citet{smk1} and used in \citet{bhardwaj14} 
to investigate possible breaks in period-color and amplitude-color relations. We plot the reddening-corrected 
magnitudes and colors against $\log(P)$ to find P-L and P-C relations. The null hypothesis is that
in the reduced model, we can fit a single regression line over the entire period range. The alternative 
hypothesis is that in the full model, we can fit two separate regression lines for stars with periods 
on opposite sides of an assumed break point $P_b$. We assume the reduced model to be the best fit under 
the null hypothesis. We write the reduced model as:

\begin{gather}
\label{eq:linearpl}
Y = a + b\log(P) ; ~~\mbox{For all $P$}. 
\end{gather}

\noindent While the full model is written as:

\begin{equation}
\label{eq:breakpl}
\begin{split}
Y = a_{S} + b_{S}\log(P) ; ~~\mbox{where $P$ $<$ $P_b$},\\
    ~~~~  = a_{L} + b_{L}\log(P) ; ~~\mbox{where $P$ $\geq$ $P_b$}.
\end{split}
\end{equation}

\noindent Here, $Y$ is the dependent variable, i.e. the magnitude when considering P-L relations, Wesenheit 
magnitudes \citep{madore82} in the case of P-W relations and the color in the case of P-C relations. The 
value of the $F$-statistic is calculated using the following equation :

\begin{gather}
F = \left(\frac{RSS_{R} - RSS_{F}}{RSS_{F}}\right) \left(\frac{\nu_{F}}{\nu_{R}-\nu_{F}}\right), 
\label{eq:fstat} 
\end{gather}

\noindent Here, RSS denotes the residual sums of squares, $\nu$ is the number of degrees of freedom, and 
subscripts $R$ and $F$ denote the reduced and full models, respectively. We compare the observed value of 
$F$ with the values obtained from the $F$-distribution under the null hypothesis at a given significance level. 
A larger value of $F$ leads to the rejection of the null hypothesis \citep[see,][]{bhardwaj14}.

We note that the parametric $F$-test is sensitive to the number of and nature of Cepheids on either side 
of the break period \citep{smk1}. Thus, samples with fewer data points, say at the long period end, or samples 
with outliers, will have a higher variance which will make it harder for the $F$-statistic to be significant.
The assumptions underlying the $F$-test are homoskedasticity, independent identically distributed observations 
and normality of residuals. \citet{bhardwaj14} have demonstrated that OGLE-III data do satisfy these assumptions 
though there are some departures by a few outliers. In order to strengthen our conclusions, we also apply a 
number of other tests that do no rely on these assumptions.

\subsection{Random walk method}

A more robust and simple approach is to study partial sums of the residuals of the least-square fits. 
The $F$-test can be affected by the distribution of the residuals from the linear fit. However, the 
non-parametric random walk test generates the distribution of the residuals from the data itself 
and, hence, provides a more robust test statistic. This test was used by \citet{koen07} to test the 
non-linearity of OGLE and MACHO $V$-band P-L relations and is outlined in brief here. 

As a first step, the data is sorted according to the increasing value of the period such that, 
$P_{1} <\dots< P_{N}$, where $N$ is the total number of stars in the sample. If $r_{k}$ are the 
residuals obtained from the linear fit to P-L relations in the form of equation \ref{eq:linearpl}, 
then the partial sum of residuals is 

\begin{gather}
C(j) = \sum_{k=1}^{j}r_{k}. 
\end{gather}

If there is no departure from the linearity, then $C(j)$ is a simple random walk. However, if there 
is a deviation from linearity $C(j)$ will not be a simple random walk. The reason is that the successive 
residuals may be correlated unlike in the case of a random walk, where the partial sums are uncorrelated 
random numbers. To quantify this, our test statistic is the vertical range of $C(j),$ i.e.

\begin{gather}
\label{eq:rvalue}
R = max[C(j)] - min[C(j)]. 
\end{gather}

If the partial sums are a random walk, $R$ will be small. Now to determine the significance level for 
the $R$-values, we use the permutation method, as the theoretical distribution of the test statistic of 
$C(j)$ is not known. We permute $r_{k}$ such that it randomizes the residuals and destroys any possible 
trends. We estimate the theoretical R value from the partial sums of permuted residuals and repeat this 
procedure for a large number ($\sim10000$) of permutations. We find the proportion of permutation $R$-values 
that exceed the observed value obtained using equation \ref{eq:rvalue}. This fraction provides 
the significance level $p(R)$, the p-value for the test statistic. $p(R)$ is the probability of the 
observed statistic under the null hypothesis (linearity); therefore, the smaller the value, the higher 
the likelihood of non-linearity.

\subsection{The testimator}
We also use the test estimator or ``testimator'' \citep{shashi07} to analyse the non-linearity of 
P-L, P-W and P-C relations at mean light. A brief overview is given here. Again, all the data are 
sorted in order of increasing period. The entire sample is then divided into $n$ different non-overlapping, 
and hence completely independent subsets. Each subset includes $N$ number of data points and the remaining 
number of data points are included in the last subset. We fit a linear regression in the form of equation 
\ref{eq:linearpl} to the first subset and determine a slope $\hat{\beta}$. This initial estimate 
(testimator) of the slope is taken as $\beta_{0}$ in the next subset, under the null hypothesis that the 
slope of the second subset is equal to the slope of the first subset, i.e. $\hat{\beta}=\beta_{0}$. 
The alternate hypothesis is that the two slopes are not equal. For hypothesis testing, we calculate 
the t-statistic, such that -

\begin{gather}
\label{eq:tvalue}
t_{obs} = \frac{(\hat{\beta} - \beta_0)}{\sqrt(MSE/S_{xx})},
\end{gather}

\noindent where,

\begin{gather*}
MSE=\frac{1}{N-2}\sum_{i=1}^{N}(Y_{i}-\hat{Y_i})^2,\\
S_{xx}=\sum_{i=1}^{N}(X_{i}-\bar{X})^2.
\end{gather*}

Here, $\hat{Y_i}$ represents the best fit linear regression and $\bar{X}$ represents the mean value of 
independent variables. Since there will be $n_{g}=n-1$ hypothesis testing in total, 
$t_{c} = t_{\alpha/2n_{g},\nu}$, and $\nu$ is the number of data points in each subset. We adopt a value 
of $\alpha=0.05$ to have a confidence level of more than $95\%$. Once we know the observed and critical 
value of the t-statistics, we can determine the coefficient $k = \left(\frac{|t_{obs}|}{t_{c}}\right)$, 
which is the probability that the initial guess of the testimator is true. If the value of $k<1$, the 
null hypothesis is accepted and we derive the new testimator slope for the next subset using the previously 
determined $\beta's$ such that -

\begin{gather}
\label{eq:betavalue}
\beta_{w} = k\hat{\beta} - (1-k)\beta_0.
\end{gather}

This value of the testimator is taken as $\beta_{0}$ for the next subset. This process of hypothesis testing 
is repeated $n_{g}$ times or untill the value $k>1$, suggesting rejection of the null hypothesis. Then we 
accept the alternate hypothesis that the data is more consistent with a non-linear relation.

\begin{table*}
\begin{minipage}{1.0\hsize}
\begin{center}
\caption{Results of the $F$-test and random walk test to discern a non-linearity at a period of 10 days on the ``simulated'' P-L relations.}
\label{table:simul_pl1}
\begin{tabular}{|l|c|c|c|c|c|c|c|c|c|c|c|c|}
\hline
\hline
$Y$  &  $b_{all}$& $\sigma_{all}$& $N_{all}$ & $b_{S}$&  $\sigma_{S}$&  $N_{S}$ &  $b_{L}$&   $\sigma_{L}$&  $N_{L}$ & $F$ &  $p(F)$&  $p(R)$ \\
\hline
             $V_{L}$&    -2.709$\pm$0.020     &     0.210&        1776&    -2.724$\pm$0.035     &     0.209&        1611&    -2.671$\pm$0.085     &     0.222&         165&     0.215&     0.806&     0.786 \\
            $V_{NL}$&    -2.759$\pm$0.020     &     0.214&        1776&    -2.984$\pm$0.035     &     0.209&        1611&    -2.541$\pm$0.085     &     0.222&         165&    31.558&     0.000&     0.000 \\
\hline
\end{tabular}
\end{center}
{\footnotesize {\bf Notes:} $b$, $\sigma$ and $N$ refer to the slope, dispersion and number of stars, respectively.
The subscripts $all,S$ and $L$ refer to the entire period range, short period range and long period range, respectively.
$p(F)$ and $p(R)$ represent the probability of acceptance of null hypothesis i.e. linear relation.}
\end{minipage}
\end{table*}

\subsection{Segmented lines and the Davies test}  
  
An alternative approach to the problem of identifying the existence of a break point is presented in 
\cite{muggeo2003}. The method first performs a linear piecewise regression considering the existence 
of the break. Thereafter, the Davies test is used to evaluate if the two segments are different enough to 
account for two  separate linear behaviours\footnote{The method is implemented in the \texttt{R} package 
\textsc{segmented} \citep{muggeo2008}.}.

The regression is performed through the use of Generalized Linear Models (GLM), a powerful regression 
tool widely used in statistics but only recently introduced in astronomy 
\citep[see e.g.][and references therein]{raichoor2012, desouza2015}. In this framework each observation 
is interpreted as a realization of a random variable and, consequently,  can be described through the 
determination of its underlying probability distribution function \citep{hilbe2012}.

Consider that the relationship between period ($X=\log(P)$, explanatory variable) and magnitudes/colours 
($Y,$ response variable) in equation~\ref{eq:breakpl} can also be described as:

\begin{gather}
\label{eq:segmented}
Y = a_{S} + b_{S}X + \Psi(X)\Delta a \left(X - X_{b}\right),
\end{gather}

\noindent where $\Delta a = a_{L} - a_{S}$, and  

\begin{gather*}
\Psi(X) = 0; ~~\mbox{if $P$ $<$ $P_b$}, \\ 
      ~~~~~~ = 1; ~~\mbox{if $P$ $\geq$ $P_b$}.
\end{gather*}

\noindent The above expression assumes, beyond the existence of the break, a continuous  transition between 
the two linear behaviours. If we want to describe a more general situation, where there is a gap in the 
intersection, the model can be written as  

\begin{gather}
\label{eq:linearpredictor}
Y = a_{S} + b_{S}X + \Psi(X)\left[\Delta a \left(X - X_b\right) - \gamma\right],
\end{gather}

\noindent where $\gamma$ represents the magnitude of the gap. The \textsc{segmented} method uses Equation 
(\ref{eq:linearpredictor}) as a linear predictor for a GLM with identity link function 
\citep[for a simple introduction to the GLM approach written for astronomers, see,][section 2]{elliott2015}.
 
The algorithm begins with an initial test point $\tilde{P_b}$ and uses the data to fit the other parameters 
in equation (\ref{eq:linearpredictor}). A new break point is updated at each iteration, 
$\hat{P}=\tilde{P} + \gamma/\Delta a$. The process is repeated until $\gamma \approx 0$, or in other words, 
when the two lines meet in the estimated break point. The uncertainty in the determination of the break is 
given by $\hat{\gamma}/\Delta a$ \citep{muggeo2008}.

Once the regression is done, our final goal is to test if the difference between the two line segments is 
evidence of distinct underlying linear behaviours. In other words, we wish to test the null hypothesis:

\begin{gather}
H_0: \Delta a(X_b) = 0,
\end{gather}

\noindent where the parameter of interest, $\Delta a$, depends on the break point, $X_b$. This is the ideal 
case for the application of the Davies test \citep{davies1987}. It assumes a $\chi^2$ distributed likelihood 
whose maximum is determined by the algorithm described above ($\hat{\Delta a}$). Given a test statistics 
($S(\bullet, X_b)$) and a list of potential break points $X_t=\{X_1, X_2, ..., X_T\}$,  Davies test states

\begin{gather}
\label{eq:pvDT}
\textrm{p-value} \geq \mathcal{N}\left(-M\right) + \frac{V\exp\left(-M^2/2\right)}{\sqrt{8\pi}},
\end{gather}

where $\mathcal{N}(\bullet)$ is the standard Normal distribution, $M=\max(|S(X_b, X_t)|)$  and $V$ is the total 
variation of the test statistics, $V=\sum_{i=2}^T|S(X_b,X_{i}) - S(X_b,X_{i-1})|$ \citep{davies1987}. 

The choice of test statistic is itself a free parameter, the only requirement being that it must follow a
normal distribution for fixed $X_b$. In the results presented here, we use the Wald statistic, which 
evaluates a test point by determining how far it is from the maximum likelihood estimates (MLE). Given the 
effect of large variances in the shape of the likelihood function, the absolute distance between test and 
MLE must be weighted by the likelihood variance. Thus, the same distance which might lead us to reject the 
test point for a tightly peaked likelihood can be judged acceptable for a broader one. In this framework, 
$S(X_b,X_t)=\hat{\Delta a}(X_b)/\sigma$, where $\hat{\Delta a}$ is the MLE  value for the difference in 
slopes and $\sigma$ is the variance of its likelihood function\footnote{It is important to emphasis that in 
this context, the algorithm will always return parameters for the two segmented lines. If the Davies test 
indicates that the break does not exist, fit parameters may be unstable for consecutive runs, even though 
the final result is not (since in this case there is no break).}.

\section{Analysis and Results}
\label{sec:analysis}

\subsection{Simulated P-L relations}

\begin{table*}
\begin{minipage}{1.0\hsize}
\begin{center}
\caption{Results of the testimator on ``simulated'' P-L relations.}
\label{table:simul_pl2}
\begin{tabular}{|c|c|c|c|c|c|c|c|c|c|c|}
\hline
\hline
Band &$n$  &  $\log(P)$&	$N$&	$\hat{\beta}$&	$\beta_{0}$&	$|t_{obs}|$&	$t_{c}$&	$k$&	Decision&	$\beta_{w}$\\
\hline
$V_{L}$&       1&    0.30223$-$0.43358   &          175&     -3.283$\pm$0.377     &        --- &        --- &        --- &        --- &        ---          &        ---\\
&       2&    0.43359$-$0.47278   &          175&     -1.416$\pm$1.208     &     -3.283&      1.546&      2.843&      0.544& Accept~$H_{0}$&     -2.268\\
&       3&    0.47279$-$0.50185   &          175&     -0.815$\pm$1.762     &     -2.268&      0.825&      2.843&      0.290& Accept~$H_{0}$&     -1.846\\
&       4&    0.50186$-$0.53508   &          175&     -1.541$\pm$1.514     &     -1.846&      0.201&      2.843&      0.071& Accept~$H_{0}$&     -1.825\\
&       5&    0.53509$-$0.57406   &          175&     -3.708$\pm$1.258     &     -1.825&      1.497&      2.843&      0.527& Accept~$H_{0}$&     -2.816\\
&       6&    0.57407$-$0.62397   &          175&     -1.941$\pm$0.921     &     -2.816&      0.950&      2.843&      0.334& Accept~$H_{0}$&     -2.524\\
&       7&    0.62398$-$0.68130   &          175&     -2.840$\pm$0.800     &     -2.524&      0.395&      2.843&      0.139& Accept~$H_{0}$&     -2.568\\
&       8&    0.68131$-$0.77611   &          175&     -2.959$\pm$0.533     &     -2.568&      0.735&      2.843&      0.258& Accept~$H_{0}$&     -2.669\\
&       9&    0.77612$-$0.94936   &          175&     -3.109$\pm$0.309     &     -2.669&      1.425&      2.843&      0.501& Accept~$H_{0}$&     -2.890\\
&      10&    0.94937$-$1.72354   &          201&     -2.758$\pm$0.058     &     -2.890&      2.293&      2.838&      0.808& Accept~$H_{0}$&     -2.783\\
\hline
$V_{NL}$&       1&    0.30223$-$0.43358   &          175&     -3.543$\pm$0.377     &        --- &        --- &        --- &        --- &        ---          &        ---\\
&       2&    0.43359$-$0.47278   &          175&     -1.675$\pm$1.208     &     -3.543&      1.546&      2.843&      0.544& Accept~$H_{0}$&     -2.527\\
&       3&    0.47279$-$0.50185   &          175&     -1.074$\pm$1.761     &     -2.527&      0.825&      2.843&      0.290& Accept~$H_{0}$&     -2.105\\
&       4&    0.50186$-$0.53508   &          175&     -1.797$\pm$1.514     &     -2.105&      0.204&      2.843&      0.072& Accept~$H_{0}$&     -2.083\\
&       5&    0.53509$-$0.57406   &          175&     -3.969$\pm$1.258     &     -2.083&      1.499&      2.843&      0.527& Accept~$H_{0}$&     -3.077\\
&       6&    0.57407$-$0.62397   &          175&     -2.201$\pm$0.921     &     -3.077&      0.952&      2.843&      0.335& Accept~$H_{0}$&     -2.784\\
&       7&    0.62398$-$0.68130   &          175&     -3.099$\pm$0.800     &     -2.784&      0.394&      2.843&      0.139& Accept~$H_{0}$&     -2.828\\
&       8&    0.68131$-$0.77611   &          175&     -3.219$\pm$0.533     &     -2.828&      0.734&      2.843&      0.258& Accept~$H_{0}$&     -2.929\\
&       9&    0.77612$-$0.94936   &          175&     -3.370$\pm$0.309     &     -2.929&      1.426&      2.843&      0.502& Accept~$H_{0}$&     -3.150\\
&      10&    0.94937$-$1.72354   &          201&     -2.543$\pm$0.058     &     -3.150&     10.426&      2.838&      3.674& Reject~$H_{0}$&     -0.919\\
\hline
\end{tabular}
\end{center}
{\footnotesize {\bf Notes:} $n$ represents the number of non-overlapping subsets and $\log(P)$ 
is the period range in each subset. $N$ and $\hat{\beta}$ represent the number of stars and 
slope of linear regression in each subset, respectively. $\beta_{0}$ and $\beta_{w}$ represent 
the initial and updated testimator after each hypothesis testing. $|t_{obs}|$ is estimated using 
equation~\ref{eq:tvalue} and $t_{c}$ represents the theoretical t-value for confidence level of 
more than 95\%. $k$ is the probability of initial guess of the testimator being true and leads 
to the decision of acceptance/rejection.}
\end{minipage}
\end{table*}

\begin{table*}
\begin{minipage}{1.0\hsize}
\begin{center}
\caption{Results of the Davies test on ``simulated'' P-L relations.}
\label{table:simul_pl3}
\begin{tabular}{|l|c|c|c|c|c|c|c|c|c|c|}
\hline
\hline
Band& N& $\log(P_b)$& $\sigma_{P_b}$& $b_{\rm low}$& $\sigma_{b_{\rm low}}$& $b_{\rm up}$& $\sigma_{b_{\rm up}}$& $c_{\rm low}$&$ c_{\rm up}$& p(D)\\ 
\hline
$V_{L}$      & 1776& 0.832& 0.366& -2.739& 0.094& -2.677& 0.098& -1.471& -1.522& 0.958\\
$V_{NL}$     & 1776& 0.868& 0.044& -3.004& 0.085& -2.448& 0.109& -1.318& -1.800& 0.000\\
\hline
\end{tabular}
\end{center}
{\footnotesize {\bf Notes:} For the Davis test, $b_{\rm low}$ refers to the slope for $P<P_b$ 
and $\sigma_{b_{\rm low}}$ is the uncertainty in the slope. Similarly, $b_{\rm up}$ refers to 
the slope for $P>P_b$ and $\sigma_{b_{\rm up}}$ is the associated uncertainty. $c_{\rm low}$ 
and $c_{\rm up}$ represent the intercept of the linear fit for $P<P_b$ and $P>P_b$, respectively. 
$p(D)$ represents the probability of acceptance of null hypothesis i.e. linear relation.}
\end{minipage}
\end{table*}

\begin{figure}
\begin{center}
\includegraphics[width=1.0\columnwidth,keepaspectratio]{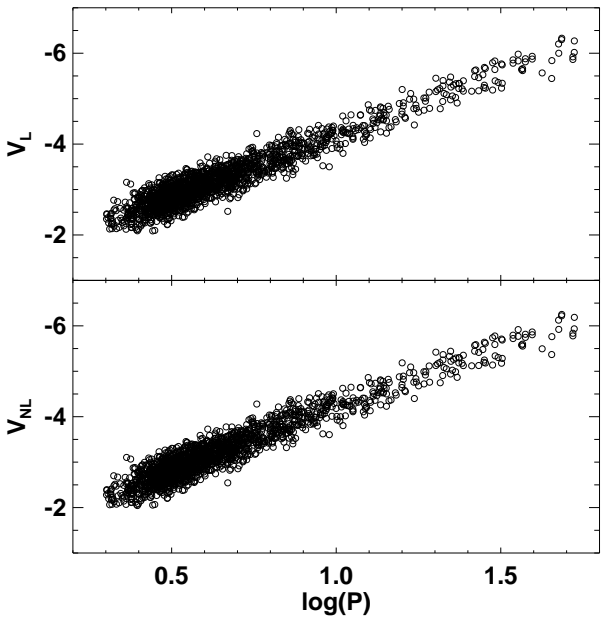}
\caption{Simulated P-L relations for OGLE-III-SS dataset. We use the equations given in 
\citet{choong06} to develop a linear ($V_{L}$) and a non-linear ($V_{NL}$) relation.}
\label{fig:simul_pl.eps}
\end{center}
\end{figure}

In order to check the accuracy of our statistical tests, we simulate two ``artificial'' 
data sets of P-L relations: one linear and one non-linear. The details of the procedure 
to simulate P-L relations are provided in \citet{choong06}. We use the equations provided 
in that paper to simulate data, but take the periods from the OGLE-III-SS sample because it 
spans a wider range and has a greater number of long-period variables. We also add the 
internal dispersion $(\sigma=0.20~mag)$ and the photometric error $(0.05~mag)$ to the 
simulated P-L relations. A plot of our simulated linear and non-linear P-L relations is 
presented in Fig.~\ref{fig:simul_pl.eps}. The two panels in this figure look very similar 
but are indeed generated differently and demonstrate why, in the era or precision cosmology, 
it is vital to determine non-linearities in P-L/P-W/P-C relations rigorously.

We use our statistical tests on these simulated data sets to determine the reliability of our 
procedures, with the results given in Table~\ref{table:simul_pl1} for the $F$-test and random 
walk, in Table~\ref{table:simul_pl2} for the testimator and in Table~\ref{table:simul_pl3} for 
the Davis test. We find that for a linear relation, the $F$-test, random walk and the Davis test 
provide a high p-value, and hence we accept the null hypothesis of a linear P-L relation. 
Similarly, the testimator accepts the null hypothesis over all period bins. Thus, these test 
results correctly predict a linear relation. When the true underlying relation is non-linear, 
the significance probabilities are zero for the $F$-test, random walk and Davis test, clearly 
allowing us to reject the null hypothesis and suggesting a non-linearity. Similarly, the 
testimator rejects the null hypothesis in a period bin that contains the break period of 10 
days. Thus our test statistics provide consistent results for the known datasets. 

\subsection{Optical-infrared P-L, P-W and P-C relations}

We apply the tests described in the previous section to various combinations of optical and 
near-infrared observations to investigate the non-linearity of P-L, P-W and P-C relations 
at mean light. We derive new P-L and P-C relations at optical wavelengths using both OGLE-III 
and OGLE-III-SS samples. We apply reddening corrections using maps taken from \citet{hasch11} 
and a reddening law from \citet{card89}. We derive the Wesenheit magnitude $W_{V,I} = V - R(V-I)$ 
for each sample, where $R=1.55$ is the color coefficient obtained using the adopted reddening 
law. We remove $2.5\sigma$ outliers to fit P-L, P-W and P-C relations at optical wavelengths. 

However, for near-infrared band data, we use the final sample of 775 FU and 474 FO stars as 
provided in Paper I, since sigma-clipping and extinction corrections were already applied in 
that work. This final sample was used to derive P-L relations in Paper I and P-W relations
in Paper II. We remove $2.5\sigma$ outliers to fit a P-C relation at these wavelengths. 
Similarly, in case of optical-infrared P-W and P-C relations, we apply reddening corrections 
and $2.5\sigma$ clipping before fitting a linear regression to these relations.
We also impose an upper limit on periods of 80 days in all these relations to avoid possible 
contamination from Ultra Long Period Cepheids \citep[ULPCs,][]{bird09}: these stars 
follow a shallower P-L relation. We do not consider FU mode Cepheids with periods below 2.5 
days because of the small sample size. 
The results of statistical tests applied to these FU 
and FO mode Cepheid datasets are discussed in following subsections.

\subsection{Fundamental-mode Cepheids}

\begin{table*}
\begin{minipage}{1.0\hsize}
\begin{center}
\caption{Results of the $F$ and random walk test on P-L, P-W and P-C relations for FU mode Cepheids 
in the LMC, to check the period breaks at 10 days. The optical band results are for OGLE-III sample. 
The $JHK_{s}$ band results are for the near-infrared counterparts of OGLE-III stars. The meaning of 
each column header is discussed in Table~\ref{table:simul_pl1}. }
\label{table:lmc_fu_10}
\scalebox{0.96}{
\begin{tabular}{|l|c|c|c|c|c|c|c|c|c|c|c|c|}
\hline
\hline
$Y$  &  $b_{all}$& $\sigma_{all}$& $N_{all}$ & $b_{S}$&  $\sigma_{S}$&  $N_{S}$ &  $b_{L}$&   $\sigma_{L}$&  $N_{L}$ & $F$ &  $p(F)$&  $p(R)$ \\
\hline
                 $V$&    -2.679$\pm$0.023     &     0.177&        1546&    -2.743$\pm$0.032     &     0.172&        1440&    -2.285$\pm$0.185     &     0.232&         106&     6.404&     0.002&     0.168\\
                 $I$&    -2.928$\pm$0.016     &     0.121&        1537&    -2.973$\pm$0.022     &     0.117&        1433&    -2.628$\pm$0.127     &     0.161&         104&     7.374&     0.001&     0.128\\
                 $J$&    -3.147$\pm$0.016     &     0.115&         774&    -3.118$\pm$0.031     &     0.109&         665&    -3.336$\pm$0.071     &     0.144&         109&     5.504&     0.004&     0.037\\
                 $H$&    -3.158$\pm$0.013     &     0.095&         774&    -3.106$\pm$0.026     &     0.090&         665&    -3.320$\pm$0.060     &     0.120&         109&     7.219&     0.001&     0.044\\
             $K_{s}$&    -3.221$\pm$0.011     &     0.084&         774&    -3.192$\pm$0.022     &     0.079&         665&    -3.361$\pm$0.055     &     0.111&         109&     5.994&     0.003&     0.076\\
\hline
           $W_{V,I}$&    -3.312$\pm$0.010     &     0.074&        1567&    -3.335$\pm$0.013     &     0.071&        1469&    -3.262$\pm$0.093     &     0.107&          98&     3.040&     0.048&     0.293\\
           $W_{J,H}$&    -3.152$\pm$0.014     &     0.107&         774&    -3.086$\pm$0.029     &     0.107&         665&    -3.297$\pm$0.055     &     0.106&         109&     6.211&     0.002&     0.085\\
       $W_{J,K_{s}}$&    -3.273$\pm$0.010     &     0.077&         774&    -3.242$\pm$0.020     &     0.074&         665&    -3.378$\pm$0.048     &     0.095&         109&     4.634&     0.010&     0.211\\
       $W_{H,K_{s}}$&    -3.361$\pm$0.013     &     0.100&         774&    -3.356$\pm$0.026     &     0.099&         665&    -3.436$\pm$0.052     &     0.107&         109&     1.240&     0.290&     0.627\\
           $W_{V,J}$&    -3.300$\pm$0.013     &     0.092&         697&    -3.289$\pm$0.025     &     0.089&         590&    -3.472$\pm$0.052     &     0.106&         107&     7.456&     0.001&     0.193\\
           $W_{V,H}$&    -3.236$\pm$0.013     &     0.094&         699&    -3.225$\pm$0.025     &     0.089&         592&    -3.393$\pm$0.058     &     0.116&         107&     5.681&     0.004&     0.158\\
       $W_{V,K_{s}}$&    -3.284$\pm$0.010     &     0.072&         698&    -3.263$\pm$0.019     &     0.068&         592&    -3.381$\pm$0.045     &     0.090&         106&     4.015&     0.018&     0.225\\
           $W_{I,J}$&    -3.288$\pm$0.016     &     0.114&         702&    -3.279$\pm$0.031     &     0.110&         593&    -3.491$\pm$0.064     &     0.127&         109&     6.719&     0.001&     0.151\\
           $W_{I,H}$&    -3.225$\pm$0.012     &     0.087&         699&    -3.184$\pm$0.024     &     0.084&         592&    -3.380$\pm$0.049     &     0.100&         107&     7.645&     0.001&     0.076\\
       $W_{I,K_{s}}$&    -3.280$\pm$0.010     &     0.075&         699&    -3.260$\pm$0.020     &     0.071&         593&    -3.378$\pm$0.046     &     0.093&         106&     3.718&     0.025&     0.320\\
         $W^H_{V,I}$&    -3.247$\pm$0.010     &     0.076&         699&    -3.220$\pm$0.020     &     0.072&         593&    -3.369$\pm$0.047     &     0.094&         106&     5.763&     0.003&     0.095\\
\hline
               $V-I$&     0.242$\pm$0.008     &     0.065&        1557&     0.218$\pm$0.012     &     0.063&        1453&     0.398$\pm$0.067     &     0.081&         104&     6.823&     0.001&     0.197\\
               $J-H$&     0.002$\pm$0.006     &     0.042&         690&    -0.012$\pm$0.011     &     0.042&         582&    -0.018$\pm$0.020     &     0.041&         108&     1.387&     0.251&     0.461\\
           $J-K_{s}$&     0.073$\pm$0.006     &     0.046&         684&     0.068$\pm$0.013     &     0.045&         577&     0.022$\pm$0.023     &     0.047&         107&     3.153&     0.053&     0.603\\
           $H-K_{s}$&     0.072$\pm$0.004     &     0.031&         688&     0.077$\pm$0.009     &     0.032&         579&     0.040$\pm$0.012     &     0.021&         109&     2.415&     0.090&     0.489\\
               $V-J$&     0.385$\pm$0.017     &     0.120&         687&     0.374$\pm$0.034     &     0.115&         578&     0.363$\pm$0.073     &     0.146&         109&     0.159&     0.853&     0.276\\
               $V-H$&     0.391$\pm$0.020     &     0.141&         686&     0.373$\pm$0.039     &     0.134&         577&     0.347$\pm$0.086     &     0.174&         109&     0.365&     0.695&     0.204\\
           $V-K_{s}$&     0.459$\pm$0.020     &     0.142&         685&     0.449$\pm$0.040     &     0.135&         577&     0.397$\pm$0.087     &     0.176&         108&     0.465&     0.628&     0.459\\
               $I-J$&     0.156$\pm$0.009     &     0.067&         690&     0.150$\pm$0.020     &     0.066&         581&     0.172$\pm$0.037     &     0.075&         109&     0.154&     0.857&     0.089\\
               $I-H$&     0.162$\pm$0.012     &     0.085&         687&     0.147$\pm$0.024     &     0.082&         578&     0.156$\pm$0.049     &     0.099&         109&     0.287&     0.751&     0.176\\
           $I-K_{s}$&     0.233$\pm$0.012     &     0.087&         687&     0.229$\pm$0.025     &     0.083&         579&     0.202$\pm$0.051     &     0.102&         108&     0.304&     0.738&     0.163\\
\hline
\end{tabular}}
\end{center}
\end{minipage}
\end{table*}

\begin{table*}
\begin{minipage}{1.0\hsize}
\begin{center}
\caption{Results from Davies test on P-L, P-W and P-C relations for FU mode Cepheids. 
The meaning of each column header is discussed in Table~\ref{table:simul_pl3}.}
\label{table:dav_fu}
\scalebox{0.96}{
\begin{tabular}{|l|c|c|c|c|c|c|c|c|c|c|}
\hline
\hline
& N& $\log(P_b)$& $\sigma_{P_b}$& $b_{\rm low}$& $\sigma_{b_{\rm low}}$& $b_{\rm up}$& $\sigma_{b_{\rm up}}$& $c_{\rm low}$&$ c_{\rm up}$& $p(D)$\\ 
\hline
$V$		&1546& 1.004& 0.076& -2.748& 0.064& -2.323& 0.292& 17.30& 16.87& 0.006\\
$I$		&1537& 1.014& 0.069& -2.977& 0.044& -2.655& 0.204& 16.76& 16.43& 0.003\\
$J$		& 774& 1.262& 0.089& -3.087& 0.044& -3.473& 0.266& 16.31& 16.79& 0.003\\
$H$		& 774& 1.259& 0.075& -3.093& 0.036& -3.469& 0.217& 15.95& 16.42& 0.000\\
$K_{s}$		& 774& 1.275& 0.081& -3.177& 0.032& -3.496& 0.204& 15.96& 16.37& 0.001\\
\hline
$W_{V,I}$	&1567& 0.899& 0.101& -3.339& 0.032& -3.231& 0.087& 15.92& 15.82& 0.098\\
$W_{J,H}$	& 774& 1.256& 0.086& -3.103& 0.039& -3.465& 0.236& 15.37& 15.83& 0.001\\
$W_{J,K_{s}}$	& 774& 1.275& 0.088& -3.237& 0.028& -3.502& 0.185& 15.73& 16.06& 0.002\\
$W_{H,K_{s}}$	& 774& 1.314& 0.138& -3.335& 0.036& -3.543& 0.251& 15.98& 16.26& 0.292\\
$W_{V,J}$	& 697& 1.261& 0.096& -3.263& 0.034& -3.537& 0.200& 15.96& 16.30& 0.005\\
$W_{V,H}$	& 699& 1.253& 0.107& -3.201& 0.036& -3.462& 0.214& 16.09& 16.42& 0.012\\
$W_{V,K_{s}}$	& 698& 1.247& 0.109& -3.255& 0.028& -3.446& 0.159& 15.82& 16.05& 0.016\\
$W_{I,J}$	& 702& 1.313& 0.085& -3.246& 0.041& -3.633& 0.286& 15.98& 16.49& 0.009\\
$W_{I,H}$	& 699& 1.250& 0.081& -3.177& 0.032& -3.494& 0.191& 15.66& 16.06& 0.000\\
$W_{I,K_{s}}$	& 699& 1.251& 0.113& -3.251& 0.028& -3.446& 0.169& 15.80& 16.05& 0.020\\
$W^{H}_{V,I}$	& 699& 1.248& 0.095& -3.214& 0.028& -3.443& 0.165& 15.76& 16.05& 0.004\\
\hline
$V-I$		&1557& 1.038& 0.069& 0.218& 0.022& 0.400& 0.127& 0.554& 0.365& 0.005\\
$J-H$		& 698& 0.471& 0.035& 0.248& 0.434& 0.001& 0.012& 0.243& 0.360& 0.453\\
$J-K_{s}$	& 684& 0.466& 0.034& 0.378& 0.540& 0.076& 0.013& 0.209& 0.350& 0.484\\
$H-K_{s}$	& 688& 1.208& 0.141& 0.083& 0.012& 0.026& 0.055&-0.014& 0.054& 0.087\\
$V-J$		& 687& 0.516& 0.049& 0.761& 0.603& 0.355& 0.041& 0.805& 1.015& 0.552\\
$V-H$		& 686& 0.523& 0.054& 0.784& 0.644& 0.361& 0.047& 1.150& 1.371& 0.554\\
$V-K_{s}$	& 685& 0.463& 0.029& 1.613& 1.882& 0.443& 0.043& 0.810& 1.352& 0.618\\
$I-J$		& 690& 0.540& 0.075& 0.279& 0.262& 0.135& 0.025& 0.376& 0.454& 0.868\\
$I-H$		& 687& 0.529& 0.058& 0.373& 0.353& 0.143& 0.029& 0.685& 0.807& 0.556\\
$I-K_{s}$	& 687& 0.481& 0.038& 0.694& 0.776& 0.222& 0.027& 0.564& 0.791& 0.592\\
\hline
\end{tabular}}
\end{center}
\end{minipage}
\end{table*}

\begin{figure*}
\begin{center}
\includegraphics[width=0.99\textwidth,keepaspectratio]{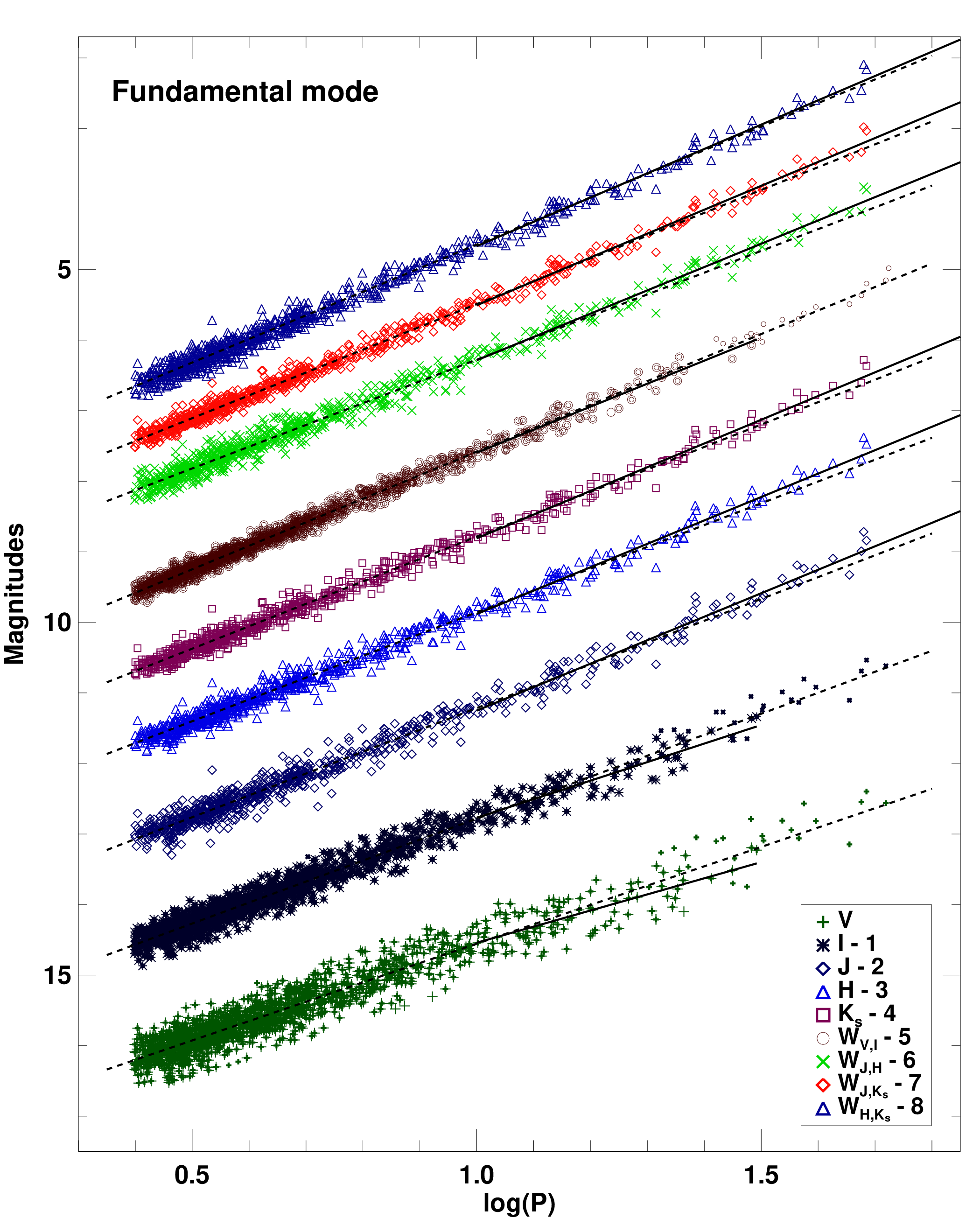}
\caption{ Optical and near-infrared P-L and P-W relations for LMC FU mode Cepheids. The dashed/solid 
lines represent the best fit regression for Cepheids with periods below and above 10 days. The smaller 
symbol for longer period Cepheids at optical bands represents stars from OGLE-III-SS.}
\label{fig:fu_plpw}
\end{center}
\end{figure*}

\begin{figure*}
\begin{center}
\includegraphics[width=0.96\textwidth,keepaspectratio]{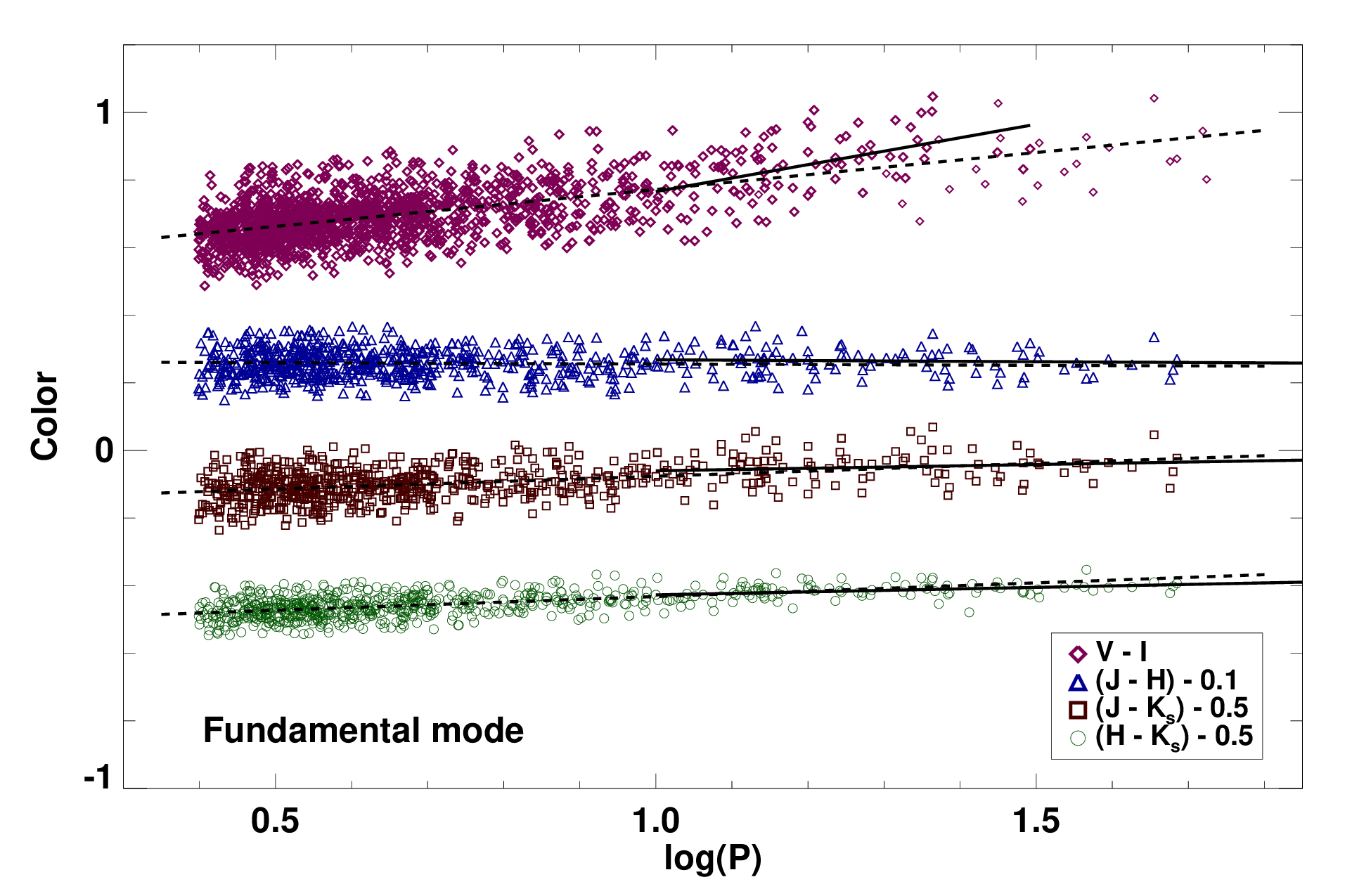}
\caption{ Optical and near-infrared P-C relations for LMC FU mode Cepheids. The dashed/solid lines 
represent the best fit regression for Cepheids with periods below and above 10 days. The smaller 
symbol for longer period Cepheids at optical bands represents stars from OGLE-III-SS.}
\label{fig:all_pc_lmc_fu.eps}
\end{center}
\end{figure*}

\begin{table*}
\begin{minipage}{1.0\hsize}
\begin{center}
\caption{Results of the $F$ and random walk test on optical bands P-L, P-W and P-C relations for FU 
mode Cepheids in OGLE-III-SS sample, to check period breaks at 10 days. The meaning of each column 
header is discussed in Table~\ref{table:simul_pl1}.}
\label{table:lmc_fu_ss}
\begin{tabular}{|l|c|c|c|c|c|c|c|c|c|c|c|c|}
\hline
\hline
$Y$  &  $b_{all}$& $\sigma_{all}$& $N_{all}$ & $b_{S}$&  $\sigma_{S}$&  $N_{S}$ &  $b_{L}$&   $\sigma_{L}$&  $N_{L}$ & $F$ &  $p(F)$&  $p(R)$ \\
\hline
                 $V$&    -2.733$\pm$0.021     &     0.179&        1569&    -2.764$\pm$0.032     &     0.172&        1441&    -2.829$\pm$0.123     &     0.251&         128&     1.577&     0.207&     0.449\\
                 $I$&    -2.964$\pm$0.014     &     0.122&        1560&    -2.979$\pm$0.022     &     0.118&        1436&    -3.020$\pm$0.083     &     0.168&         124&     0.893&     0.409&     0.450\\
           $W_{V,I}$&    -3.330$\pm$0.009     &     0.074&        1590&    -3.335$\pm$0.013     &     0.071&        1470&    -3.407$\pm$0.051     &     0.105&         120&     2.746&     0.064&     0.448\\
               $V-I$&     0.228$\pm$0.007     &     0.065&        1580&     0.218$\pm$0.012     &     0.063&        1453&     0.194$\pm$0.042     &     0.087&         127&     1.285&     0.277&     0.675\\
\hline
\end{tabular}
\end{center}
\end{minipage}
\end{table*}

\begin{table*}
\begin{minipage}{1.0\hsize}
\caption{Results from Davies test on P-L, P-W and P-C relations for Cepheids in the 
OGLE-III-SS sample. The meaning of each column header is discussed in Table~\ref{table:simul_pl3}.}
\label{table:dt_ss}
\begin{center}
\begin{tabular}{|l|c|c|c|c|c|c|c|c|c|c|}
\hline
\hline
Band& N& $\log(P_b)$& $\sigma_{P_b}$& $b_{\rm low}$& $\sigma_{b_{\rm low}}$& $b_{\rm up}$& $\sigma_{b_{\rm up}}$& $c_{\rm low}$& $c_{\rm up}$& $p(D)$\\ 
\hline
$V$      & 1569& 1.308& 0.144& -2.711& 0.048& -3.069& 0.514& 17.28&  17.750& 0.494\\
$I$      & 1560& 1.354& 0.143& -2.950& 0.032& -3.253& 0.460& 16.740& 17.150& 0.484\\
$W_{V,I}$& 1590& 1.249& 0.159& -3.318& 0.021& -3.453& 0.171& 15.910& 16.070& 0.081\\
$V-I$    & 1580&   1.369&  0.129&  0.237&  0.017&  0.029&  0.264&  0.543&  0.827&  0.129\\
\hline
\end{tabular}
\end{center}
\end{minipage}
\end{table*}

In \citet{bhardwaj14}, we found a significant change in the slope of the P-C relation at 
maximum and minimum light for FU mode Cepheids with periods greater than 10 days in the LMC. 
The non-linearity in the optical band P-L and P-C relations at a period of 10 days for Cepheid 
variables is discussed extensively in previous studies (see references in \S1). \citet{ngeow09} 
used $F$-test to determine non-linearity in P-L relations for FU mode Cepheids using OGLE-III data. 
We revisit the breaks at 10 days in P-L relation observed by \citet{ngeow09} with more 
test-statistics and determine their statistical significance in order to compare with our new 
near-infrared data. The results of the $F$-test and the random walk method for FU mode Cepheids at 
multiple bands are presented in Table~\ref{table:lmc_fu_10}. The results of the testimator and 
the Davis test are given in Table~\ref{table:testimator_fu_1} and Table~\ref{table:dav_fu}, respectively. 

The optical and near-infrared band P-L and P-W relations for FU mode Cepheids are presented in
Fig.~\ref{fig:fu_plpw}. Visual inspection of the $V$- and $I$-band P-L relations 
reveals a small variation in the slope. The results of test statistics are as follows: 

\begin{itemize}
\item{The optical band P-L relations are non-linear at 10 days according to the $F$-test, testimator and the Davis test.}
 
\item{Since the optical band P-L relations have higher dispersion with consequently larger residuals at longer periods using a 
single regression line, the random walk does not provide evidence to support a non-linearity at the 
$90\textrm{-}95\%$ significance level. The random walk gives a p-value of $0.168$ and $0.128$ for $V$- and 
$I$-band P-L relations, respectively.}

\item{As the $F$-test, Davis test and the testimator result in a non-linear relation, we consider the $V$- and $I$-band 
P-L relations to be non-linear.}

\item {The $F$-test and the Davis test also indicate that the optical-band P-W relation is nonlinear, although the 
deviation from a single slope is considerably smaller.}

\item{Similarly, the P-C relation shown in Fig.~\ref{fig:all_pc_lmc_fu.eps}, exhibit a significant non-linearity 
at 10 days in the optical bands.}
\end{itemize}

\begin{figure}
\begin{center}
\includegraphics[width=1.0\columnwidth,keepaspectratio]{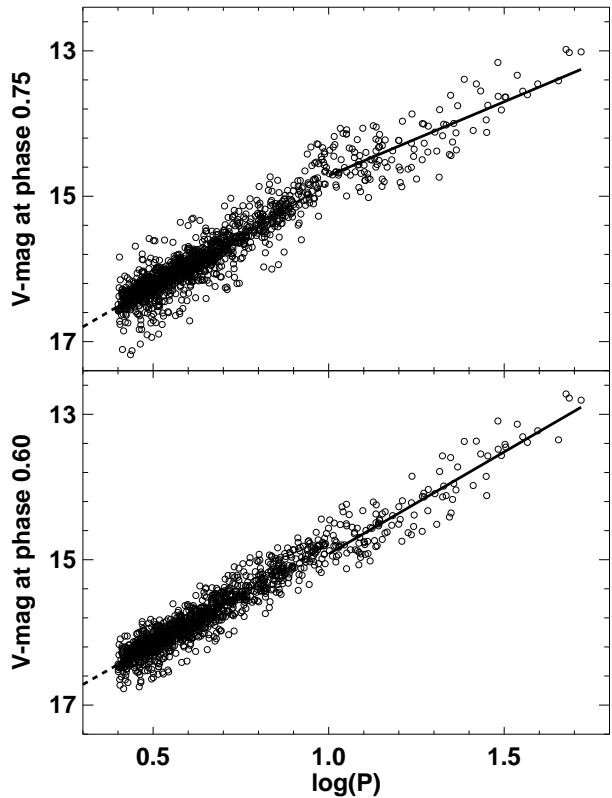}
\caption{The $V$-band P-L relations for LMC FU mode Cepheids using OGLE-III-SS data. The dashed/solid 
lines represent the best fit regression for Cepheids with periods below and above 10 days.}
\label{fig:pl_ss}
\end{center}
\end{figure}

We note, however, that these results do not hold if we use the OGLE-III-SS data set. The 
results for the $F$-test and random walk are presented in Table~\ref{table:lmc_fu_ss}, while 
Table~\ref{table:ss_testimator} and Table~\ref{table:dt_ss} present the results of the testimator 
and the Davis test, respectively. All the test statistics rule out the possibility of a break at 
10 days in the P-L relations. We note that while adding the longer-period variables increases the 
sample size relative to OGLE-III, it also results in increased dispersion for longer period range 
P-L relations. Similarly, the P-C relation using OGLE-III-SS data also does not provide any evidence 
of a non-linearity at 10 days, according to all the test statistics. Since we observed a non-linearity 
in the P-C relations at maximum and minimum light for this sample in \citet{bhardwaj14}, we plan 
to look into the characteristics of the multi-phase P-L and P-C relations before reaching any 
further conclusions. We present the $V$-band P-L relation at two phases of pulsation using 
OGLE-III-SS data set, displayed in Fig.~\ref{fig:pl_ss}. All the test statistics provide evidence 
of a significantly non-linear P-L relation at these phases. A detailed statistical analysis of these 
relation as a function of phase will be presented in a subsequent study. However, the P-W relation 
for the OGLE-III-SS sample does provide evidence of a marginally significant non-linearity using the 
$F$-test (p-value $\sim0.06$), the Davis test (p-value $\sim0.08$) and the testimator. We note from 
Table~\ref{table:lmc_fu_ss} that the dispersion in this P-W relation is less (only $0.074$ mag), 
compared to $0.179$ and $0.122$ for the $V$- and $I$-band P-L relations.

We extend our analysis to test for the first time the possibility of non-linearities in the near-infrared 
P-L and P-W relations derived using time-series observations. The results of the test statistics are as follows:

\begin{itemize}

\item{A small but statistically significant change in the slope for longer period Cepheids is observed in the 
near-infrared P-L relations, according to all four test statistics.}

\item{The Davis test suggests a break around $\log(P)\sim1.25\textrm{-}1.30$ in the near-infrared P-L relations. Similarly,
\citet{anupam15} observed a distinct variation in the Fourier amplitude parameters for longer wavelengths as compared to optical
bands around a period of 20 days. This indicates that the observed non-linearities in these relations are 
in fact associated with sharp variations in light curve structure as discussed in \citet{bhardwaj14}. It is a
very interesting result relating changes in light curve parameters with breaks in P-L and/or P-C relations, 
i.e. physical parameters.}

\item{In the case of the near-infrared P-W relations, $W_{J,H}$ and $W_{J,K_s}$ are 
found to be non-linear at $\log(P)\sim1.25$, according to most test statistics.} 

\item{However, $W_{H,K_s}$ is linear according to the $F$, random walk and Davis tests. As the longer wavelengths 
are expected to be less sensitive to metallicity and extinction, this linearity may be very useful in distance 
scale applications. Further, this may also indicate the effect of metallicities on the linearity of a P-L relation 
and will be subsequently investigated as a function of pulsation phase.}

\item{We also note that the $W^H_{V,I}$ relation, which is commonly used \citep{riess09, riess11} for calibrating 
the distances to SNe Ia host galaxies and determining the value of the Hubble constant, $H_0$, is non-linear 
according to all four test statistics. Again, the Davis test provides evidence of a significant change in the slope 
at $\log(P)\sim1.248$.}

\item{The optical-near-infrared P-W relations are also non-linear according to most test statistics, resulting in 
the adopted non-linear relation for all combinations.}

\item{The P-C relation at near-infrared wavelengths are shown in Fig.~\ref{fig:all_pc_lmc_fu.eps} together with the 
optical band P-C relation. However, these relations do not provide any evidence of possible non-linearity at 10 days and 
suggest that the temperature fluctuations are less dominant at these wavelengths.} 

\item{We note that the multi-wavelength LMC P-L relations are found to be non-linear in this study but the SMC P-L 
relations are linear at 10 days \citep{ngeow15}.}

\end{itemize}

\subsection{First-overtone mode Cepheids}

Non-linearity in the P-L and P-C relations is discussed in detail in the literature for FU mode 
Cepheids. In \citet{bhardwaj14}, we also found significant breaks in the P-C and A-C relations at 
2.5 days for FO mode Cepheids in the Magellanic Clouds at the phases of maximum and minimum light. 
We extend that work to see if these breaks exist in the P-L, P-W and P-C relations at mean light in 
multiple bands. The results of the $F$-test and the random-walk method are summarized in 
Table~\ref{table:lmc_fo_2.5}, while Table~\ref{table:testimator_fo_1} and Table~\ref{table:dav_fo} 
present the results for the testimator and the Davis test, respectively.

\begin{table*}
\begin{center}
\caption{Results for $F$ test and random walk on P-L, P-W and P-C relations for FO mode Cepheids in LMC, to 
check period breaks at 2.5 days. The meaning of each column header is discussed in Table~\ref{table:simul_pl1}.}
\label{table:lmc_fo_2.5}
\scalebox{0.96}{
\begin{tabular}{|l|c|c|c|c|c|c|c|c|c|c|c|c|}
\hline
\hline
$Y$  &  $b_{all}$& $\sigma_{all}$& $N_{all}$ & $b_{S}$&  $\sigma_{S}$&  $N_{S}$ &  $b_{L}$&   $\sigma_{L}$&  $N_{L}$ & $F$ &  $p(F)$&  $p(R)$ \\
\hline
                 $V$&    -3.299$\pm$0.029     &     0.182&        1084&    -3.410$\pm$0.045     &     0.187&         802&    -2.688$\pm$0.108     &     0.155&         282&    16.080&     0.000&     0.000\\
                 $I$&    -3.354$\pm$0.021     &     0.129&        1084&    -3.430$\pm$0.032     &     0.132&         802&    -2.894$\pm$0.078     &     0.112&         282&    17.231&     0.000&     0.000\\
                 $J$&    -3.336$\pm$0.038     &     0.114&         474&    -3.455$\pm$0.064     &     0.117&         348&    -3.107$\pm$0.103     &     0.102&         126&     3.166&     0.043&     0.122\\
                 $H$&    -3.258$\pm$0.029     &     0.092&         474&    -3.322$\pm$0.049     &     0.097&         348&    -3.117$\pm$0.075     &     0.075&         126&     2.000&     0.136&     0.284\\
             $K_{s}$&    -3.276$\pm$0.025     &     0.081&         474&    -3.303$\pm$0.042     &     0.087&         348&    -3.179$\pm$0.062     &     0.062&         126&     0.979&     0.376&     0.277\\
\hline
           $W_{V,I}$&    -3.466$\pm$0.011     &     0.069&        1075&    -3.517$\pm$0.017     &     0.070&         790&    -3.359$\pm$0.045     &     0.066&         285&     9.145&     0.000&     0.000\\
           $W_{J,H}$&    -3.076$\pm$0.035     &     0.119&         474&    -3.099$\pm$0.061     &     0.130&         348&    -3.141$\pm$0.093     &     0.084&         126&     0.388&     0.678&     0.505\\
       $W_{J,K_{s}}$&    -3.216$\pm$0.024     &     0.082&         474&    -3.194$\pm$0.042     &     0.089&         348&    -3.230$\pm$0.060     &     0.062&         126&     0.269&     0.764&     0.324\\
       $W_{H,K_{s}}$&    -3.319$\pm$0.035     &     0.119&         474&    -3.264$\pm$0.060     &     0.130&         348&    -3.295$\pm$0.088     &     0.089&         126&     0.939&     0.392&     0.026\\
           $W_{V,J}$&    -3.436$\pm$0.029     &     0.095&         422&    -3.526$\pm$0.049     &     0.099&         301&    -3.288$\pm$0.084     &     0.082&         121&     3.522&     0.030&     0.109\\
           $W_{V,H}$&    -3.391$\pm$0.028     &     0.093&         421&    -3.463$\pm$0.046     &     0.096&         300&    -3.241$\pm$0.082     &     0.082&         121&     2.904&     0.056&     0.179\\
       $W_{V,K_{s}}$&    -3.293$\pm$0.021     &     0.071&         421&    -3.322$\pm$0.037     &     0.078&         300&    -3.252$\pm$0.053     &     0.053&         121&     0.622&     0.537&     0.488\\
           $W_{I,J}$&    -3.433$\pm$0.036     &     0.118&         420&    -3.519$\pm$0.061     &     0.124&         299&    -3.291$\pm$0.106     &     0.101&         121&     2.052&     0.130&     0.249\\
           $W_{I,H}$&    -3.254$\pm$0.026     &     0.086&         422&    -3.325$\pm$0.046     &     0.095&         301&    -3.221$\pm$0.061     &     0.062&         121&     2.218&     0.110&     0.174\\
       $W_{I,K_{s}}$&    -3.280$\pm$0.022     &     0.074&         420&    -3.299$\pm$0.039     &     0.081&         299&    -3.249$\pm$0.055     &     0.056&         121&     0.275&     0.760&     0.404\\
         $W^H_{V,I}$&    -3.314$\pm$0.021     &     0.070&         421&    -3.394$\pm$0.037     &     0.076&         300&    -3.232$\pm$0.053     &     0.055&         121&     4.580&     0.011&     0.049\\
\hline
               $V-I$&     0.064$\pm$0.010     &     0.064&        1086&     0.033$\pm$0.016     &     0.066&         803&     0.269$\pm$0.038     &     0.054&         283&    13.344&     0.000&     0.002\\
               $J-H$&    -0.113$\pm$0.015     &     0.050&         408&    -0.139$\pm$0.026     &     0.054&         293&    -0.026$\pm$0.039     &     0.039&         115&     2.150&     0.118&     0.200\\
           $J-K_{s}$&    -0.089$\pm$0.017     &     0.057&         408&    -0.139$\pm$0.028     &     0.058&         291&     0.048$\pm$0.052     &     0.052&         117&     4.851&     0.008&     0.169\\
           $H-K_{s}$&     0.035$\pm$0.013     &     0.045&         416&     0.013$\pm$0.024     &     0.048&         295&     0.046$\pm$0.035     &     0.036&         121&     0.777&     0.460&     0.043\\
               $V-J$&     0.110$\pm$0.038     &     0.126&         407&     0.064$\pm$0.062     &     0.130&         290&     0.353$\pm$0.116     &     0.112&         117&     2.183&     0.114&     0.121\\
               $V-H$&    -0.024$\pm$0.042     &     0.142&         404&    -0.091$\pm$0.070     &     0.147&         288&     0.316$\pm$0.123     &     0.122&         116&     3.495&     0.031&     0.090\\
           $V-K_{s}$&    -0.010$\pm$0.045     &     0.152&         406&    -0.107$\pm$0.075     &     0.157&         290&     0.359$\pm$0.134     &     0.135&         116&     3.947&     0.020&     0.114\\
               $I-J$&     0.062$\pm$0.022     &     0.074&         404&     0.045$\pm$0.036     &     0.075&         285&     0.163$\pm$0.070     &     0.068&         119&     1.090&     0.337&     0.378\\
               $I-H$&    -0.083$\pm$0.026     &     0.088&         403&    -0.124$\pm$0.044     &     0.092&         287&     0.132$\pm$0.072     &     0.071&         116&     3.636&     0.027&     0.024\\
           $I-K_{s}$&    -0.072$\pm$0.028     &     0.094&         404&    -0.143$\pm$0.046     &     0.097&         288&     0.174$\pm$0.082     &     0.082&         116&     4.860&     0.008&     0.043\\
\hline
\end{tabular}}
\end{center}
\end{table*}

\begin{table*}
\caption{Results from Davies test on P-L, P-W and P-C relations for FO mode Cepheids. 
The meaning of each column header is discussed in Table~\ref{table:simul_pl3}.}
\label{table:dav_fo}
\begin{center}
\scalebox{0.97}{
\begin{tabular}{|l|c|c|c|c|c|c|c|c|c|c|}
\hline
\hline
& N& $\log(P_b)$& $\sigma_{P_b}$& $b_{\rm low}$& $\sigma_{b_{\rm low}}$& $b_{\rm up}$& $\sigma_{b_{\rm up}}$& $c_{\rm low}$& $c_{\rm up}$& $p(D)$\\ 
\hline
\hline
$V$& 		1084& 0.446& 0.030& -3.468& 0.084& -2.684& 0.314& 16.90& 16.55& 0.000\\
$I$& 		1084& 0.449& 0.030& -3.474& 0.058& -2.920& 0.229& 16.32& 16.07& 0.000\\
$J$& 		 474& 0.427& 0.064& -3.484& 0.116& -3.104& 0.290& 15.87& 15.71& 0.059\\
$H$& 		 474& 0.422& 0.075& -3.332& 0.087& -3.094& 0.212& 15.49& 15.39& 0.156\\
$K_{s}$&	 474& 0.019& 0.086& -2.860& 0.820& -3.301& 0.055& 15.46& 15.47& 0.735\\
\hline
$W_{V,I}$&     1075& 0.381& 0.039& -3.533& 0.036& -3.340& 0.082& 15.43& 15.35& 0.000\\
$W_{J,H}$& 	474& 0.149& 0.106& -3.317& 0.427& -3.026& 0.096& 14.90& 14.86& 0.510\\
$W_{J,K_{s}}$& 	474& 0.070& 0.050& -2.761& 0.468& -3.257& 0.058& 15.17& 15.21& 0.045\\
$W_{H,K_{s}}$& 	474& 0.066& 0.030& -2.247& 0.664& -3.410& 0.083& 15.38& 15.45& 0.000\\
$W_{V,J}$&	422& 0.391& 0.068& -3.522& 0.094& -3.278& 0.177& 15.49& 15.40& 0.091\\
$W_{V,H}$& 	421& 0.409& 0.074& -3.462& 0.086& -3.237& 0.188& 15.62& 15.53& 0.172\\
$W_{V,K_{s}}$& 	421& 0.014& 0.066& -2.815& 0.728& -3.314& 0.047& 15.29& 15.30& 0.261\\
$W_{I,J}$& 	420& 0.394& 0.089& -3.514& 0.116& -3.281& 0.224& 15.52& 15.43& 0.323\\
$W_{I,H}$& 	422& 0.210& 0.091& -3.406& 0.214& -3.197& 0.079& 15.19& 15.15& 0.257\\
$W_{I,K_{s}}$& 	420& 0.072& 0.058& -2.929& 0.409& -3.313& 0.055& 15.27& 15.29& 0.134\\
$W^{H_{V,I}}$& 	421& 0.373& 0.060& -3.386& 0.071& -3.201& 0.114& 15.29& 15.22& 0.046\\
\hline
$V-I$&	       1086& 0.473& 0.032&  0.014& 0.028& 0.298& 0.134& 0.581& 0.447& 0.000\\
$J-H$& 		408& 0.417& 0.077& -0.147& 0.048&-0.022& 0.111& 0.373& 0.321& 0.216\\
$J-K_{s}$& 	408& 0.460& 0.042& -0.157& 0.046& 0.124& 0.153& 0.398& 0.269& 0.000\\
$H-K_{s}$& 	416& 0.151& 0.044& -0.171& 0.159& 0.078& 0.035& 0.033&-0.005& 0.002\\
$V-J$& 		407& 0.465& 0.099&  0.047& 0.110& 0.350& 0.402& 1.049& 0.908& 0.432\\
$V-H$& 		404& 0.455& 0.079& -0.123& 0.121& 0.274& 0.411& 1.425& 1.244& 0.191\\
$V-K_{s}$& 	406& 0.456& 0.065& -0.136& 0.127& 0.385& 0.440& 1.449& 1.212& 0.043\\
$I-J$& 		404& 0.280& 0.186&  0.091& 0.124& 0.021& 0.083& 0.448& 0.468& 1.000\\
$I-H$& 		403& 0.455& 0.092& -0.134& 0.074& 0.074& 0.252& 0.837& 0.742& 0.384\\
$I-K_{s}$& 	404& 0.453& 0.060& -0.157& 0.080& 0.186& 0.269& 0.864& 0.709& 0.024\\
\hline
\end{tabular}}
\end{center}
\end{table*}

\begin{figure*}
\begin{center}
\includegraphics[width=0.99\textwidth,keepaspectratio]{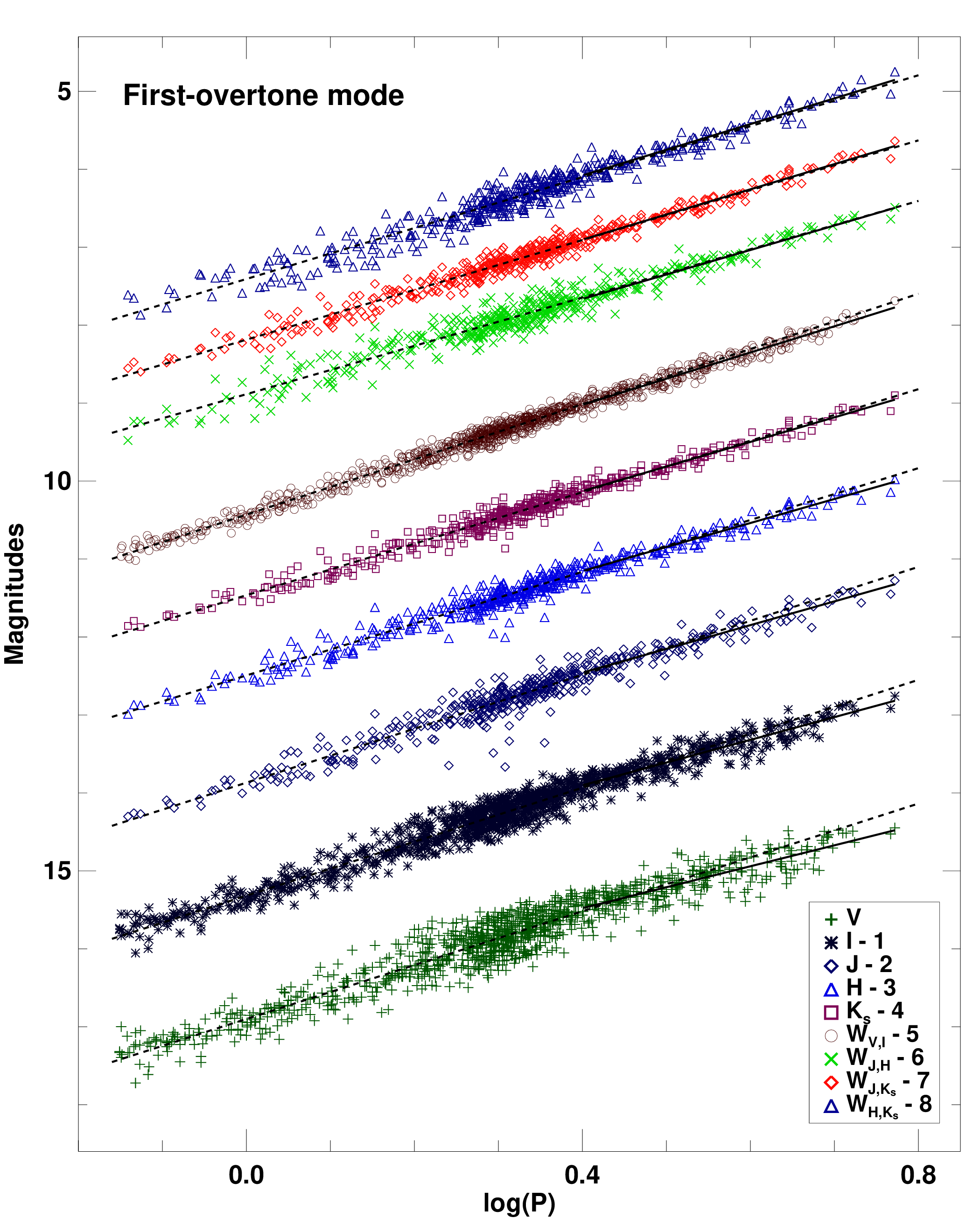}
\caption{ Optical and near-infrared P-L and P-W relations for LMC FO mode Cepheids. The dashed/solid 
lines represent the best fit regression for Cepheids with periods below and above 2.5 days.}
\label{fig:fo_plpw}
\end{center}
\end{figure*}

\begin{figure*}
\begin{center}
\includegraphics[width=0.96\textwidth,keepaspectratio]{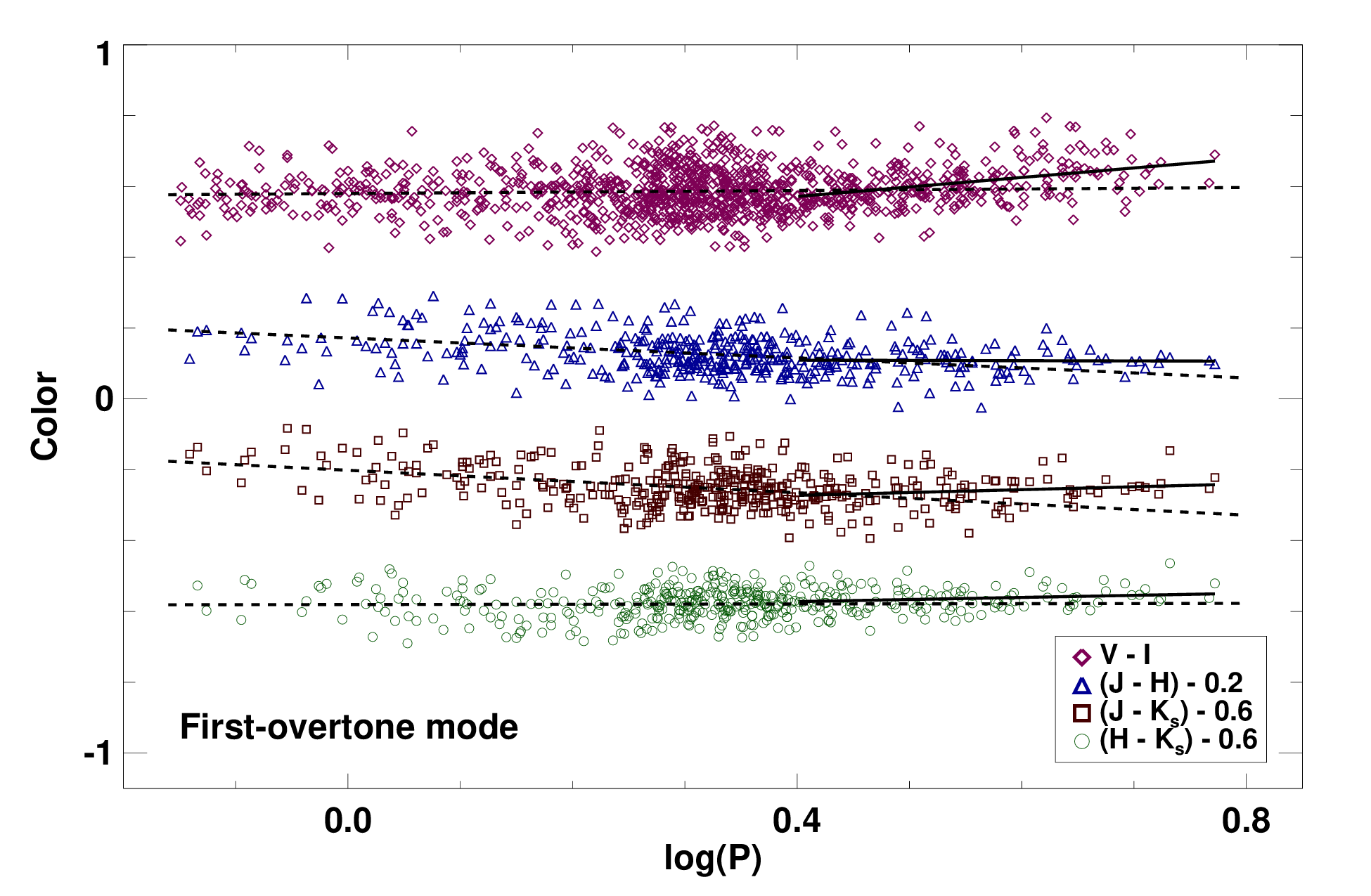}
\caption{ Optical and near-infrared P-C relations for LMC FO mode Cepheids. The dashed/solid lines 
represent the best fit regression for Cepheids with periods below and above 10 days.}
\label{fig:all_pc_lmc_fo.eps}
\end{center}
\end{figure*}

\begin{table*}
\begin{center}
\caption{Results of non-linearities in P-L, P-W and P-C relations for FU and 
FO mode Cepheids in LMC at multiple wavelengths.} 
\label{table:final_results}
\scalebox{1.0}{
\begin{tabular}{|p{1.3cm}|c|c|c|c|p{0.8cm}|c|c|c|c|c|}
\hline
\hline
\multicolumn{11}{c}{OGLE-III + CPAPIR}\\
\hline
			&  \multicolumn{5}{c}{\bf FU}			&  \multicolumn{5}{c}{\bf FO} \\
\hline
			&FT	&RW	&TM	&DT	&AD		&FT	&RW	&TM	&DT	&AD\\
\hline
  $V_{L}$	&\ding{56} &\ding{56} &\ding{56} &\ding{56} &\ding{56} 		&--	&--	&--	&--	&--	\\
 $V_{NL}$	&\ding{52} &\ding{52} &\ding{52} &\ding{52} &\ding{52}		&--	&--	&--	&--	&--	\\
\hline
   $V$		&\ding{52} &\ding{56} &\ding{52} &\ding{52} &\ding{52}	&\ding{52} &\ding{52} &\ding{52} &\ding{52} &\ding{52}\\
   $I$		&\ding{52} &\ding{56} &\ding{52} &\ding{52} &\ding{52}	&\ding{52} &\ding{52} &\ding{52} &\ding{52} &\ding{52}\\ 
   $J$		&\ding{52} &\ding{52} &\ding{52} &\ding{52} &\ding{52}	&\ding{52} &\ding{56} &\ding{52} &\ding{56} &\ding{52}\\
   $H$		&\ding{52} &\ding{52} &\ding{52} &\ding{52} &\ding{52}	&\ding{56} &\ding{56} &\ding{52} &\ding{56} &\ding{56}\\
 $K_{s}$	&\ding{52} &\ding{52} &\ding{52} &\ding{52} &\ding{52}	&\ding{56} &\ding{56} &\ding{52} &\ding{56} &\ding{56}\\
\hline
$W_{V,I}$	&\ding{52} &\ding{56} &\ding{52} &\ding{56} &\ding{52}	&\ding{52} &\ding{52} &\ding{52} &\ding{52} &\ding{52}\\
$W_{J,H}$    	&\ding{52} &\ding{52} &\ding{52} &\ding{52} &\ding{52}	&\ding{56} &\ding{56} &\ding{52} &\ding{56} &\ding{56}\\
$W_{J,K_{s}}$	&\ding{52} &\ding{56} &\ding{52} &\ding{52} &\ding{52}	&\ding{56} &\ding{56} &\ding{56} &\ding{56} &\ding{56}\\
$W_{H,K_{s}}$	&\ding{56} &\ding{56} &\ding{52} &\ding{56} &\ding{56}	&\ding{56} &\ding{52} &\ding{56} &\ding{56} &\ding{56}\\
$W_{V,J}$	&\ding{52} &\ding{56} &\ding{56} &\ding{52} &\ding{52}	&\ding{52} &\ding{56} &\ding{52} &\ding{56} &\ding{52}\\
$W_{V,H}$	&\ding{52} &\ding{56} &\ding{56} &\ding{52} &\ding{52}	&\ding{56} &\ding{56} &\ding{52} &\ding{56} &\ding{56}\\
$W_{V,K_{s}}$	&\ding{52} &\ding{56} &\ding{52} &\ding{52} &\ding{52}	&\ding{56} &\ding{56} &\ding{56} &\ding{56} &\ding{56}\\
$W_{I,J}$	&\ding{52} &\ding{56} &\ding{56} &\ding{52} &\ding{52}	&\ding{56} &\ding{56} &\ding{56} &\ding{56} &\ding{56}\\
$W_{I,H}$	&\ding{52} &\ding{52} &\ding{56} &\ding{52} &\ding{52}	&\ding{56} &\ding{56} &\ding{56} &\ding{56} &\ding{56}\\
$W_{I,K_{s}}$	&\ding{52} &\ding{56} &\ding{52} &\ding{52} &\ding{52}	&\ding{56} &\ding{56} &\ding{56} &\ding{56} &\ding{56}\\
$W^H_{V,I}$	&\ding{52} &\ding{52} &\ding{52} &\ding{52} &\ding{52}	&\ding{52} &\ding{52} &\ding{52} &\ding{52} &\ding{52}\\
\hline
$V-I$		&\ding{52} &\ding{56} &\ding{56} &\ding{52} &\ding{52}	&\ding{52} &\ding{52} &\ding{52} &\ding{52} &\ding{52}\\
$J-H$		&\ding{56} &\ding{56} &\ding{56} &\ding{56} &\ding{56}	&\ding{56} &\ding{56} &\ding{56} &\ding{56} &\ding{56}\\
$J-K_{s}$	&\ding{56} &\ding{56} &\ding{56} &\ding{56} &\ding{56}	&\ding{52} &\ding{56} &\ding{52} &\ding{52} &\ding{52}\\
$H-K_{s}$	&\ding{56} &\ding{56} &\ding{56} &\ding{56} &\ding{56}	&\ding{56} &\ding{52} &\ding{56} &\ding{56} &\ding{56}\\
$V-J$		&\ding{56} &\ding{56} &\ding{52} &\ding{56} &\ding{56}	&\ding{56} &\ding{56} &\ding{52} &\ding{56} &\ding{56}\\
$V-H$		&\ding{56} &\ding{56} &\ding{56} &\ding{56} &\ding{56}	&\ding{52} &\ding{52} &\ding{52} &\ding{56} &\ding{52}\\
$V-K_{s}$	&\ding{56} &\ding{56} &\ding{56} &\ding{56} &\ding{56}	&\ding{52} &\ding{56} &\ding{52} &\ding{52} &\ding{52}\\
$I-J$		&\ding{56} &\ding{52} &\ding{56} &\ding{56} &\ding{56}	&\ding{56} &\ding{56} &\ding{56} &\ding{56} &\ding{56}\\
$I-H$		&\ding{56} &\ding{56} &\ding{56} &\ding{56} &\ding{56}	&\ding{52} &\ding{52} &\ding{52} &\ding{56} &\ding{52}\\
$I-K_{s}$	&\ding{56} &\ding{56} &\ding{56} &\ding{56} &\ding{56}	&\ding{52} &\ding{52} &\ding{52} &\ding{52} &\ding{52}\\
\hline
\multicolumn{11}{c}{OGLE-III-SS}\\
\hline
   $V$    	&\ding{56} &\ding{56}	&\ding{56} &\ding{56}	&\ding{56}	        &-- 	&--	&--	&--	&--\\
   $I$	    	&\ding{56} &\ding{56}	&\ding{56} &\ding{56}	&\ding{56}	        &--	&--	&--	&--	&--\\
$W_{V,I}$    	&\ding{56} &\ding{56}	&\ding{52} &\ding{56}	&\ding{56}	        &--	&--	&--	&--	&--\\
$V-I$	    	&\ding{56} &\ding{56}	&\ding{56} &\ding{56}	&\ding{56}	        &--	&--	&--	&--	&--\\
\hline
\hline
\hline
\end{tabular}}
\end{center}

{\footnotesize 
{\begin{enumerate}
\item For each band P-L, P-W or P-C relations, \ding{52}/\ding{56} represent the break/no break under each test statistics.\\
\item For the $F$-test, the break is accepted if $F$-value $>$ 3.\\ 
\item For the random walk, it is accepted if $p(R)<0.10$. \\
\item For the testimator, we accept a break if null hypothesis is rejected for the subset which includes break period.\\
\item For the Davies test, we accept a break if $p(D)<0.05$ (equation \ref{eq:pvDT}).
\item The adopted result is listed under ``AD'' column. The break is accepted if atleast two tests result in a `\ding{52}'. 
\end{enumerate}
}}
\end{table*}

The optical and near-infrared band P-L and P-W relations for FO mode Cepheids are presented in 
Fig.~\ref{fig:fo_plpw}. Both optical band P-L relations present evidence of a clear non-linearity 
around 2.5 days based on all four test statistics. The P-W relation is also consistent with a 
break at the same period using all test statistics. It is important to note that the Davis test 
suggests a break at $\log(P) = 0.45~(P\sim2.8)$~days in P-L relations. Similarly, the testimator also provides
evidence of a break in period bin covering range from 2.5 days to 5 days. A further trend can be seen 
in optical P-L relations at 0.6 days but there are not enough data points so we discard any such 
non-linearity and these stars are not shown in Fig.~\ref{fig:fo_plpw}.

At near-infrared wavelengths, we find a small but significant change in the slope of the $J$-band 
P-L relation using the $F$ and the testimator tests, but the random walk and Davis test do not find 
evidence to support a non-linear relation. However, all the tests except the testimator, provide 
evidence of a linear P-L relation in $H$- \& $K_s$. So for FO mode stars, evidence for a 
non-linearity diminishes with increasing wavelength. Similarly, in the case of P-W relations, the 
$F$-test, Davis test and the testimator results in a linear relation in $W_{H,K_s}$. All four test 
statistics result in a non-linear $W^H_{V,I}$ relation. Therefore, both FU and FO mode Cepheid follow 
a linear P-W relation in $W_{H,K_s}$ but a non-linear relation in $W^H_{V,I}$. The optical and 
near-infrared combinations of P-W relations provide mixed results depending on the test statistics 
under consideration and result in adopted linear relations for most of the combinations. The optical 
and near-infrared bands P-C relations for FO mode Cepheids are shown in Fig.~\ref{fig:all_pc_lmc_fo.eps}. 
All tests yield significant detection of non-linearity at 2.5 days in optical bands. At near-infrared 
wavelengths, only $(J-K_s)$ color provide evidence of a change in slope of P-C relation. The mixed 
optical-infrared band colors result in a non-linear P-C relation for most combinations. The test 
results for all FU and FO mode Cepheids are summarized in Table~\ref{table:final_results}.

\subsection{Robustness of the test-statistics}

\begin{figure}
\begin{center}
\includegraphics[width=0.47\textwidth,keepaspectratio]{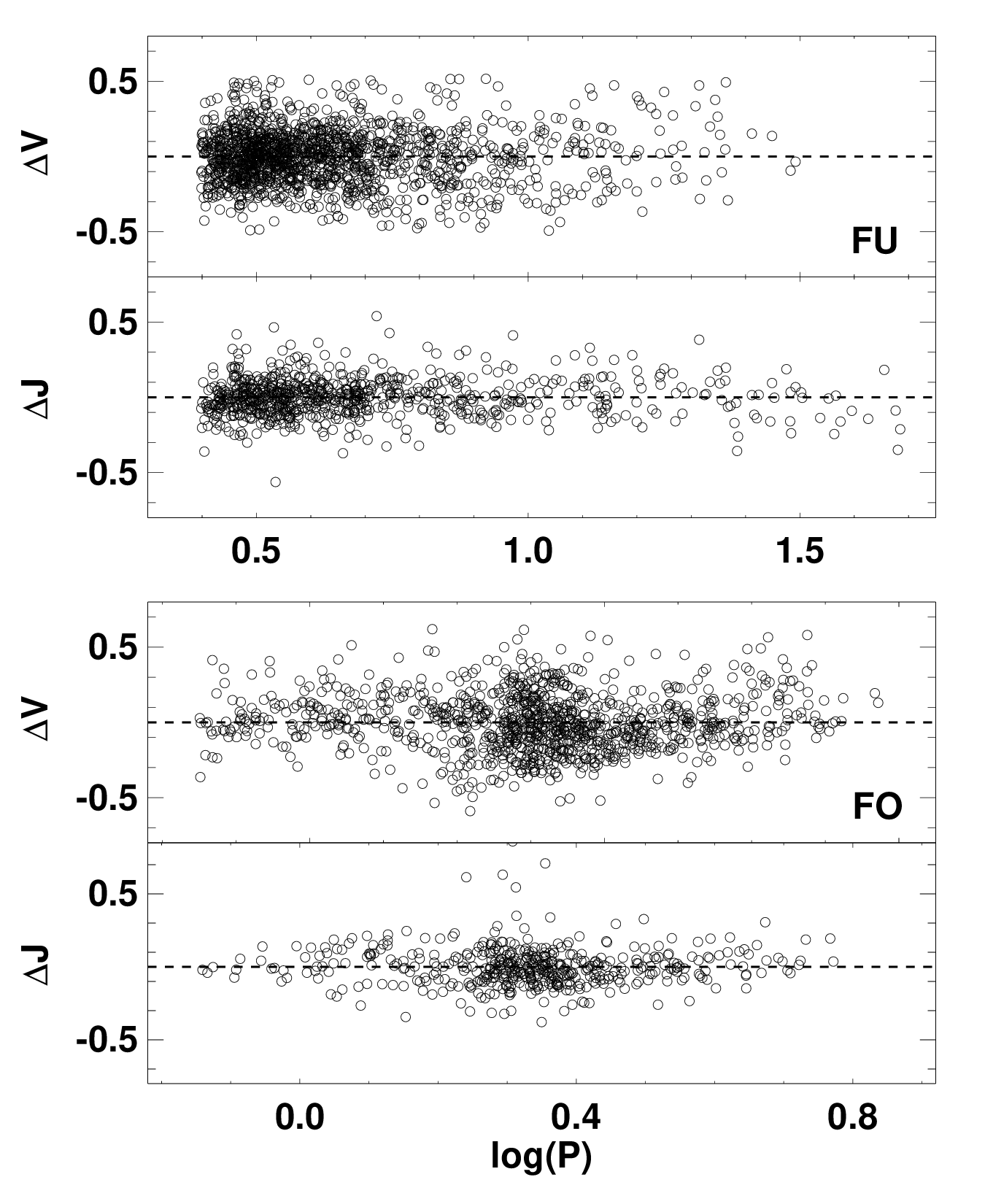}
\caption{The residuals of a linear regression to $V$- and $J$-band P-L relation for FU and FO mode Cepheids in the LMC.}
\label{fig:residuals.eps}
\end{center}
\end{figure}

\begin{table}
\begin{minipage}{1.0\hsize}
\begin{center}
\caption{Results of $F$-test using permutation and bootstrap methods.} 
\label{table:boot_permu}
\begin{tabular}{|l|c|c|c|c|c|c|}
\hline
\hline
			&  \multicolumn{3}{c}{\bf FU}			&  \multicolumn{3}{c}{\bf FO} \\
\hline
Band		    &	$F_0$	&$p_b(F)$	&$p_p(F)$&	$F_0$	&$p_b(F)$	&$p_p(F)$\\
\hline
 \multicolumn{7}{c}{OGLE-III}\\
\hline
                  $V$&      6.404&      0.001&      0.002&     16.080&      0.000&      0.000\\
                 $I$&      7.374&      0.001&      0.001&     17.231&      0.000&      0.000\\
                 $J$&      5.504&      0.004&      0.004&      3.166&      0.044&      0.040\\
                 $H$&      7.219&      0.001&      0.001&      2.000&      0.135&      0.140\\
             $K_{s}$&      5.994&      0.003&      0.004&      0.979&      0.383&      0.377\\
\hline
           $W_{V,I}$&      3.040&      0.049&      0.052&      9.145&      0.000&      0.000\\
           $W_{J,H}$&      6.211&      0.002&      0.002&      0.388&      0.683&      0.679\\
       $W_{J,K_{s}}$&      4.634&      0.010&      0.011&      0.269&      0.759&      0.768\\
       $W_{H,K_{s}}$&      1.240&      0.287&      0.286&      0.939&      0.396&      0.381\\
           $W_{V,J}$&      7.456&      0.000&      0.001&      3.522&      0.028&      0.027\\
           $W_{V,H}$&      5.681&      0.003&      0.004&      2.904&      0.056&      0.054\\
       $W_{V,K_{s}}$&      4.015&      0.019&      0.018&      0.622&      0.539&      0.550\\
           $W_{I,J}$&      6.719&      0.002&      0.001&      2.052&      0.129&      0.125\\
           $W_{I,H}$&      7.645&      0.000&      0.001&      2.218&      0.113&      0.114\\
       $W_{I,K_{s}}$&      3.718&      0.023&      0.027&      0.275&      0.758&      0.762\\
         $W^H_{V,I}$&      5.763&      0.003&      0.004&      4.580&      0.011&      0.010\\
\hline
               $V-I$&      6.823&      0.001&      0.001&     13.344&      0.000&      0.000\\
               $J-H$&      1.387&      0.250&      0.251&      2.150&      0.119&      0.113\\
           $J-K_{s}$&      3.153&      0.042&      0.043&      4.851&      0.008&      0.008\\
           $H-K_{s}$&      2.415&      0.090&      0.090&      0.777&      0.460&      0.466\\
               $V-J$&      0.159&      0.853&      0.856&      2.183&      0.115&      0.115\\
               $V-H$&      0.365&      0.696&      0.700&      3.495&      0.032&      0.031\\
           $V-K_{s}$&      0.465&      0.632&      0.624&      3.947&      0.018&      0.021\\
               $I-J$&      0.154&      0.856&      0.857&      1.090&      0.338&      0.333\\
               $I-H$&      0.287&      0.750&      0.752&      3.636&      0.025&      0.028\\
           $I-K_{s}$&      0.304&      0.742&      0.745&      4.860&      0.009&      0.008\\
\hline
\multicolumn{7}{c}{OGLE-III-SS}\\
\hline
                 $V$&      1.577&      0.203&      0.208 &--- &--- &--- \\
                 $I$&      0.893&      0.423&      0.404 &--- &--- &--- \\
           $W_{V,I}$&      2.746&      0.065&      0.065 &--- &--- &--- \\
               $V-I$&      1.285&      0.273&      0.271 &--- &--- &--- \\
\hline
\end{tabular}
\end{center}
\end{minipage}
\end{table}

As discussed in Section~\ref{sec:method}, the $F$-distribution requires the independent and 
identically distributed random variables and works under the assumption of homoskedasticity 
and normality of residuals. In order to validate our $F$-test results, we need to investigate if 
our data follows these assumptions. Since observations of each Cepheid are independent of other 
observations, the assumption of variables being independent and identically distributed is satisfied. 
Fig.~\ref{fig:residuals.eps} displays the residuals of linear regression to $V$- and $J$-band 
P-L relations for FU and FO mode Cepheids in the LMC. We do not see any significant trend in 
the residuals, that can violate the assumption of homoskedasticity. Further, the mean residuals 
for all these relations are consistent with zero and therefore, the homoskedasticity assumption 
holds well for our data. 

\begin{figure*}
\centering
  \begin{tabular}{@{}cc@{}}
    \includegraphics[width=.45\textwidth]{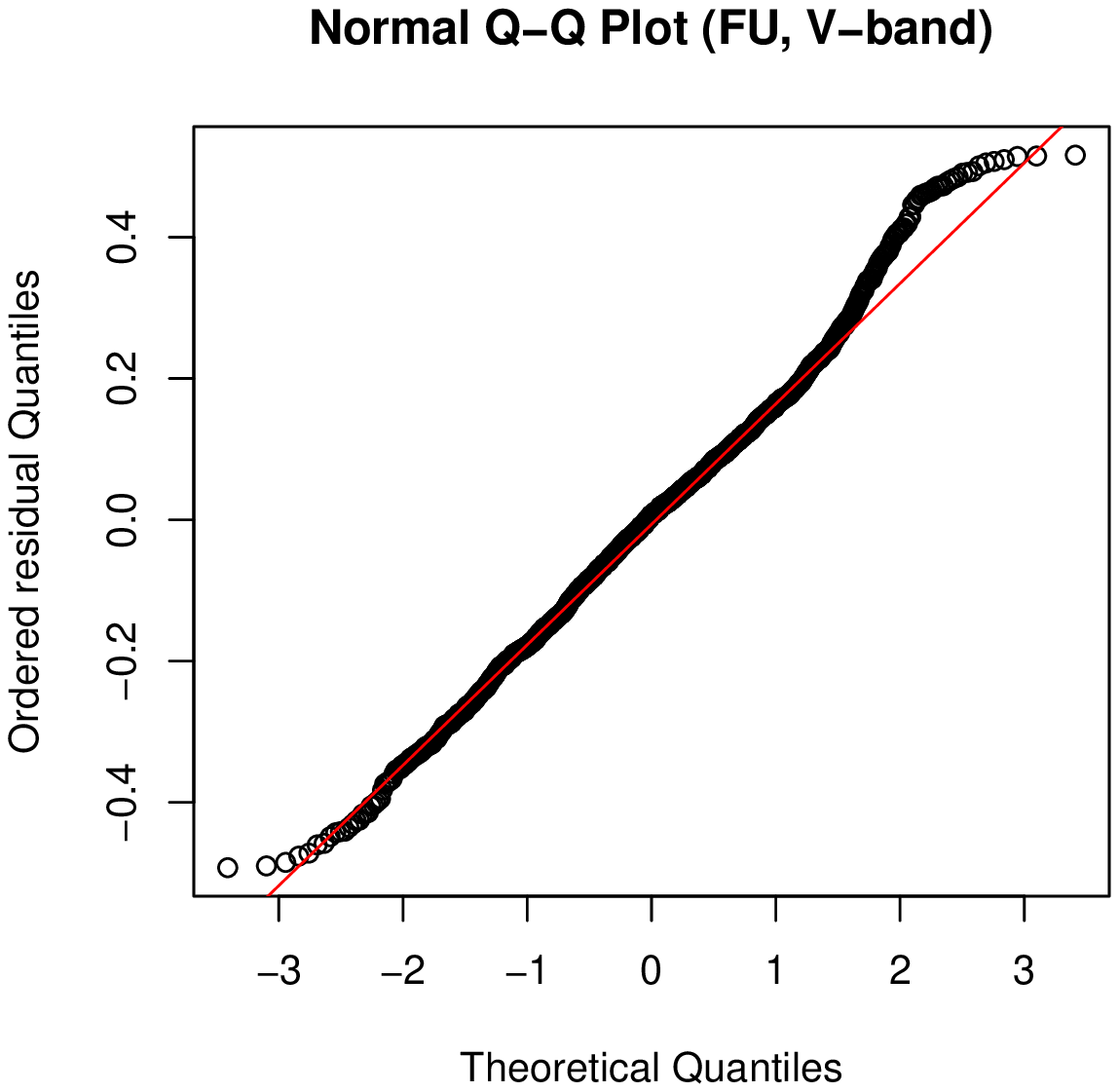} &
    \includegraphics[width=.45\textwidth]{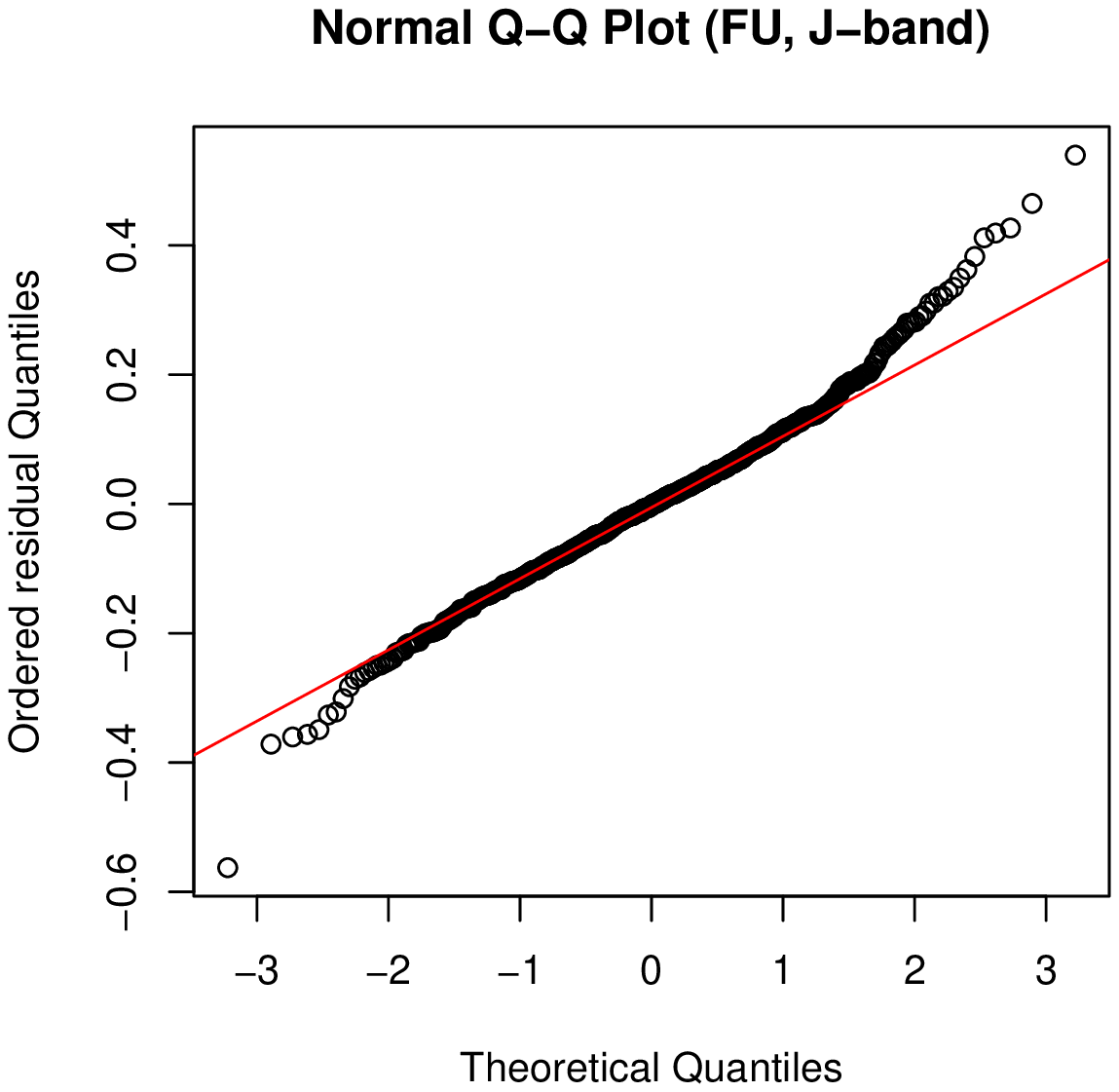}   \\
    \includegraphics[width=.45\textwidth]{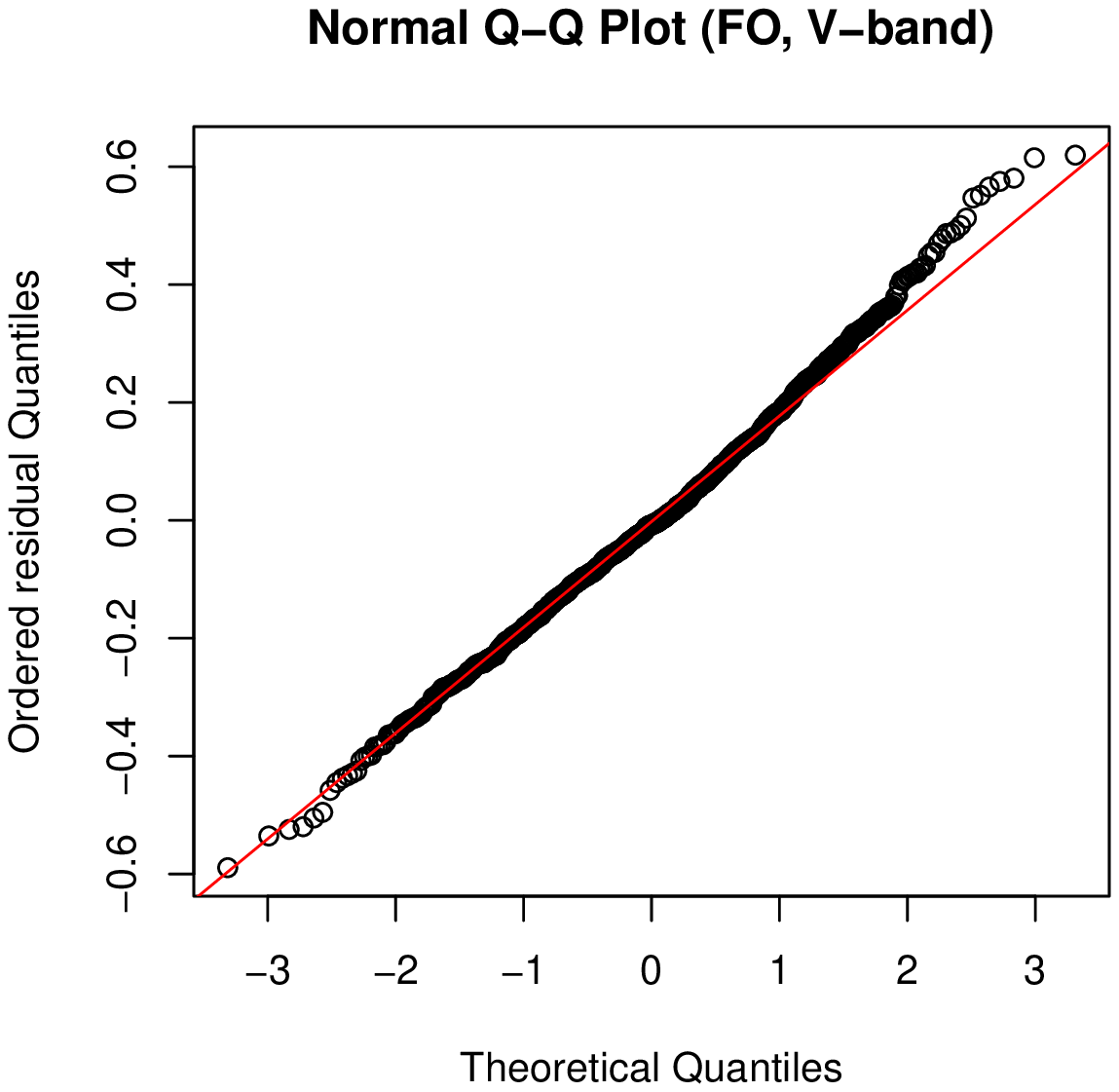} &
    \includegraphics[width=.45\textwidth]{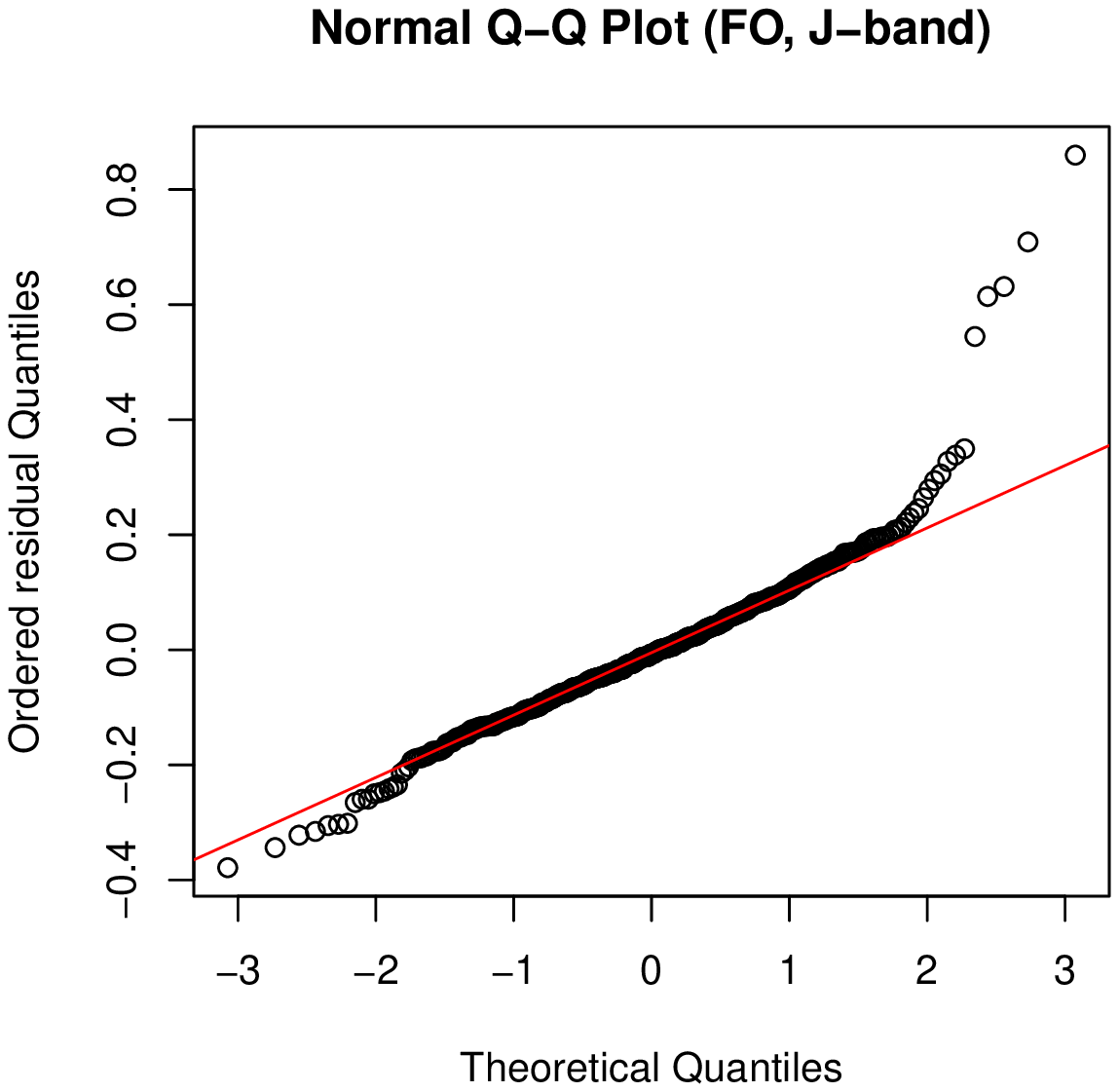}   \\
  \end{tabular}
 \caption{The quantile-quantile plot for the residuals of $V$- and $J$-band P-L relation for FU and FO mode Cepheids.}
 \label{fig:qqplots.eps}
\end{figure*}

However, we see some trend on the extreme ends in the quantile-quantile (q-q) plots for the P-L, 
P-W and P-C relations. These plots are generally used to describe the normality of residuals. We 
have displayed the q-q plot for $V$- and $J$-band P-L relations for FU and FO mode Cepheids in 
Fig.~\ref{fig:qqplots.eps}. Even though the majority of residual quantiles fall on the $y=x$ line, 
there are some outliers on the extreme ends. In order to ensure that these outliers do not belong to 
only long or short period Cepheids, we locate these outliers on the P-L plots and find that they 
are spread over the entire period range. Since the normality of the residuals is essential to 
trust the $F$-test results under theoretical $F$-distribution, we re-do our analysis without these
outliers and find consistent results. Moreover, we used the following two methods to generate a 
theoretical $F$-distribution that is independent of the way residuals are distributed.

\subsubsection{Permutation method}

This method is similar to the random walk test. At first, we fit a single regression line and a two line fit 
with a break at 10 days and estimate the observed value of $F$-statistics ($F_{0}$) using equation~\ref{eq:fstat}. 
In order to develop theoretical $F$-statistic, we take the residuals of a single regression line and re-sample 
them without replacement. It means that each residual is considered in the new sample but at a different 
location selected at random. We add the new sample of residuals to the best fit linear regression to generate a 
pseudo dataset. Again from equation~\ref{eq:fstat}, we get a $F$-value ($F_{i}$) for the pseudo dataset. 
This procedure is repeated $\sim10000$ times. The proportion of the $F_{i}$, in the pseudo dataset, that 
are greater than the observed $F_{0}$ in the actual dataset, gives the p-value ($p_p(F)$) i.e. the 
probability that underlying relation is linear. 

\subsubsection{Bootstrap method}

In this method, we follow the same procedure as in permutation method except that the residuals are re-sampled
with replacement. This means that it is not necessary that each residual from single regression line is considered
in the new sample. This subset consists of randomly selected residuals out of the original residuals, where each 
selected residual may appear more than once. Again, this sample is added to best fit linear regression to generate 
pseudo data and determine the $F$-statistic. The steps are repeated $\sim10000$ times to obtain the p-value ($p_b(F)$).

The results of the permutation and bootstrap methods are listed in Table~\ref{table:boot_permu}. The probabilities, 
$p_p(F)$ and $p_b(F)$ are found to be almost identical to $p(F)$ values listed in Table~\ref{table:lmc_fu_10} and 
\ref{table:lmc_fo_2.5} for FU and FO mode Cepheids. This confirms that $F$-test results are preserved even with slight 
departure of residuals from normality in q-q plots.

We also test the reliability of our statistical tests with different internal dispersions around 
the underlying relation, whether it be linear or non-linear. We find that both the $F$-test and random 
walk methods provide correct results upto a dispersion limit of $\sigma \sim 0.5$. Further, the random 
walk results are also sensitive to the size of residuals. Hence, this test provides the most accurate 
result for a relation with least internal dispersion. The testimator is highly sensitive to the number 
of data points in each subset and their dispersion. After significant experimentation, we find that a 
nearly equal number ($\sim200)$ of stars in each subset with 3 minimum steps of hypothesis testing is 
the best choice that provide accurate result even for a relatively greater dispersion. For a value of 
$\sigma > 0.3$, the results of the testimator do not hold as the dispersion of P-L in each subset increases
significantly due to a small period range. However, the results of the Davis test hold well for even 
greater internal dispersions as compared to other tests. We also note that the dispersion for short
and long periods range P-L relations are different and hence it is important to introduce different 
scatter for various period ranges and simultaneously test our results. Even though the data becomes 
rarer for long period Cepheids, our results hold well as most of the P-L, P-W and P-C relations have 
$\sigma < 0.2$. A detailed study on new tests with different dispersions for different period
bins will be presented in a subsequent communication.

\subsection{Constraints from other Cepheid hosts}
\label{sec:imp}

The {\it Supernovae and $H_0$ for the Equation of State of dark energy} (SH0ES) project \citep[][hereafter, R09,
R11]{riess09, riess11} determined $H_0$ with a total uncertainty of $\sim~5~\&~3\%$, respectively, using optical and
near-infrared observations of Cepheids in host galaxies of type Ia supernovae (SNe Ia). The Cepheid distances to these
host galaxies are estimated using the calibrated P-L relations for Cepheids in the Milky Way, LMC and NGC$\,$4258. We
note that most of the Cepheids observed in the SNe host galaxies have $P>10$~days. Given the long-term goal of reducing
the uncertainty in $H_0$ to $\sim1\%$, it is important to characterize the effect of possible non-linear P-L relations 
on the distance scale.

In this analysis, we adopt the methodology of R09 and R11 to determine the effect of linear and non-linear P-L 
relations derived from a global fit to all Cepheid data, as presented in R11.  Briefly, the Wesenheit magnitudes 
for the $j^{th}$ Cepheid in the $i^{th}$ SN host is defined as follows:

\begin{eqnarray}
\label{cep}
m_{w_{i,j}} & = & \mu_{0,i} + M_w + b_{w}(\log P_{i,j}~\textrm{-}~1) + Z_{w}\Delta\log[O/H]_{i,j}, \nonumber  \\
\end{eqnarray}

\noindent where $m_{w_{i,j}}$ are the $W^H_{V,I}$ magnitudes derived in Paper II, $\mu_{0,i}$ are the 
distances to the SNe hosts, $M_w$ is the calibrated Wesenheit magnitude for the Cepheid at 10 days in 
the calibrator galaxy, $b_{w}$ is the (presumably universal) slope of the P-L relation and 
$Z_{w}$ represents the metallicity dependence. The last three terms represent the absolute Wesenheit magnitudes
for Cepheids.
All four test statistics suggest that this P-W relation is non-linear and the Davis test suggests a break 
around 18 days. Simultaneously, for the SNe host, the $V$-band peak magnitude 
(corrected for extinction, light curve shape, host galaxy mass, etc.) in the $i^{th}$ host 
galaxy is expressed as:

\begin{equation}
\label{sn}
m^0_{{v,i}} = \mu_{0,i} - M^0_V.
\end{equation}

We include SNe magnitudes in our analysis to provide a tighter constraint on the global fit and also to 
compare with R11 results. $M^0_V$ represents the calibrated peak magnitude of a SN Ia in the calibrator 
galaxy. We combine equations~\ref{cep} \& \ref{sn} into matrix form $y=Lq$, similar to equation~18 in 
R09 and solve using $\chi^2$ minimization. We use the $m^0_{v,i}$ for the SNe Ia from the values listed 
in Table 3 of R11. The matrix implementation requires a covariance matrix $C$ of measurement errors that 
includes photometric uncertainties in the Cepheid magnitudes and uncertainties in the SNe template fits, 
are adopted from R11. We estimate the values of the free parameters and their uncertainties using 
equation~5 from R09. We calculate the value of $\chi^2_{dof}$ and use it to re-scale the covariance 
matrix of errors.  

We consider the following cases based on different choices of calibrator(s): (a) three galaxies (NGC$\,$4258 + LMC + MW)
(b) LMC only; (c) NGC$\,$4258 only. We use new LMC distance of $18.493\pm0.0476$~mag \citep{piet13} and adopt the revised 
distance modulus for NGC$\,$4258 of $29.404\pm0.065$~mag \citep{hump13} to calibrate P-W relations in these galaxies. 
The results of the global fits for these three cases are presented in Table~\ref{table:hc}. Note that cases 
b \& c only differ in the value and uncertainty of $M_w$. Column 1 represents the following variants considered 
in our analysis:

\begin{itemize}
\item {\bf1}  - The initial fit based on a sample of 510 Cepheids from R11 (using only the LMC 
Cepheids from \citet[][hereafter, P04]{persson04}). 
\item {\bf2}  - Replacing P04 LMC Cepheids in variant `1' with 710 Cepheids having $W^H_{V,I}$ magnitudes from Paper II.
\item {\bf3}  - Finally, restricting the LMC sample in variant `2' to only Cepheids with $P>10$~days.
\end{itemize}  

In order to address a particular case, we will refer to the calibrator choice followed by the variant (e.g., a1
refers to the three-calibrator case with variant `1'). Our slopes are shallower for cases a1 \& b1 and steeper for c1
relative to R11 (who obtained -3.21, -3.19 and -3.02, respectively), perhaps because we did not adopt a prior on the
slope of the Milky Way or the LMC P-L relations.  Cases a2, b2 and c2 show that the adoption of the new LMC sample
results in a significant improvement of the constraint in the global slope, and the resulting value is in fact
identical to the three-calibrator case of R11. It also provides a better constraint on the value of the metallicity
coefficient. Lastly, cases a3, b3 and c3 show that restricting the LMC sample to long-period Cepheids only
does not make any significant difference to the slope.

\begin{table}
\scalebox{0.95}{
\begin{minipage}{1.0\hsize}
\begin{center}
\caption{Results of the global fit for $W^H_{V,I}$ relation using Cepheids in SNe host galaxies.} 
\label{table:hc}
\begin{tabular}{lrcccccc}
\hline
\hline
 &  $N$& $b_w$& 	$\sigma(b_w$)& 	$M_w$& 		$\sigma(M_w)$&	$Z_w$& 	$\sigma(Z_w)$ \\
\hline
\multicolumn{8}{c}{(a) Three calibrators}\\
\hline
1&  510&     -3.117&      0.038&     -5.957&      0.027&     -0.050&      0.070\\
2& 1165&     -3.212&      0.013&     -5.944&      0.019&      0.054&      0.054\\
3&  564&     -3.233&      0.037&     -6.015&      0.026&      0.030&      0.066\\
\hline
\multicolumn{8}{c}{(b) LMC as single calibrator} \\
\hline
1&  510&     -3.092&      0.040&     -6.029&      0.053&     -0.200&      0.093\\
2& 1165&     -3.217&      0.013&     -6.062&      0.039&     -0.163&      0.074\\
3&  564&     -3.196&      0.039&     -6.114&      0.050&     -0.169&      0.090\\
\hline
\multicolumn{8}{c}{(c) NGC$\,$4258 as single calibrator} \\
\hline
1&  510&     -3.092&      0.040&     -5.954&      0.032&     -0.200&      0.093\\
2& 1165&     -3.217&      0.013&     -5.908&      0.023&     -0.163&      0.074\\
3&  564&     -3.196&      0.039&     -6.010&      0.031&     -0.169&      0.090\\
\hline
\end{tabular}
\end{center}
\end{minipage}}
\vspace{5pt}\\
{\footnotesize {\bf Notes:} $N$ represents the number of stars. $b_w$, $M_w$ and $Z_w$ refer to the slope, zero-point
and metallicity coefficient of the global fit solution, respectively. The corresponding columns with $\sigma$ 
represent the uncertainties in these parameters.}
\end{table}    
 
We recall from Table~\ref{table:lmc_fu_10} that the slopes of the $W^H_{V,I}$ relations for all periods, short periods,
and long periods are $b_{all} = -3.247\pm0.010$, $b_{S} = -3.220\pm0.020$ and $b_{L} = -3.369\pm0.047$,
respectively. From Table~\ref{table:hc}, we note that the slopes of the global-fit solutions for the linear and
non-linear versions (variants `2' and `3') are very consistent with the LMC-based values for $b_{all}$ and $b_S$, but
significantly different from $b_L$. In fact, the slope of the global-fit solution with LMC calibration is identical 
to short-period version of the solution based exclusively on LMC data.

The slopes obtained from a global fit to Cepheids in SH0ES galaxies using the linear and non-linear versions of
the LMC P-L relations are very similar, independent of the choice of the calibrator galaxy. Therefore, we do not expect
any significant impact of this parameter on the distance scale or the value of $H_0$, considering that the dispersion of
the {\it HST}-based P-L relations ($\sim0.3\textrm{-}0.4$~mag) is more than three times the dispersion of the LMC P-L
relation. Therefore, even a significant change in slope for the long-period P-W relation will be masked by the dispersion
of the global fit and will not have any impact on distance parameters until more precise relations can be obtained with
{\it James Webb Space Telescope}. However, our LMC P-W sample provides a better constraint on the slope of the global fit,
 leading to more accurate distance measurements and improved constraints on the metallicity dependence.

\section{Discussion and Conclusions}
\label{sec:discuss}

In this work, we studied possible non-linearities in the Period-Luminosity, Period-Wesenheit $\&$ 
Period-Color relations for fundamental and first-overtone mode Cepheids in the LMC at optical and 
near-infrared wavelengths by applying the robust statistical tests. We also determine the 
significance of these observed non-linearities in distance scale applications. The results are 
summarized as follows-

\begin{enumerate}

\item For fundamental-mode Cepheids, we find that optical band P-L relation and Wesenheit function 
are non-linear at 10 days. At near-infrared wavelengths, the P-L relations are non-linear around 18
days. It is suggested that the break around 18 days is related to a distinct separation in mean Fourier 
amplitude parameters around 20 days for longer wavelengths as compared to optical bands. We also observe 
small but significant evidence of non-linearity in most of the P-W relations. The P-C relations exhibit 
non-linearity only at the optical wavelengths. 

\item For first-overtone mode Cepheids, we find a significant non-linearity at 2.5 days for all relations 
in the optical bands. These results are in good agreement with breaks at maximum and minimum light P-C 
relations observed in \citet{bhardwaj14}. Most of the P-L and P-W relations based on a combination of 
optical and near-infrared magnitudes are linear, while most P-C relations are non-linear.

\item The $W_{H,K_s}$ relation for both fundamental and first-overtone mode Cepheid is perfectly 
linear according to most test statistics. Since the longer wavelengths are less sensitive to 
extinction and metallicity, this linearity may be useful for future distance scale applications.

\item The $W^H_{V,I}$ relation used by the SH0ES project is found to be non-linear according to 
all test statistics. The Davis test suggest a break around $\log(P)=1.25$. We re-analyse the 
SH0ES data and obtain a global slope of $\textrm{-}3.212\pm0.013$ based on three calibrators. We 
note that this slope is consistent with the value derived from LMC variables when considering only 
short periods or the entire period range, while it is significantly different from the value 
derived from long-period LMC variables.

\item We do not find any significant difference in the slope derived from the global fit using 
linear or non-linear P-L relations, due to the large observational scatter in the distant Cepheid 
data. Our LMC sample does provide a better constraint on the slope and on the metallicity coefficient 
of the global-fit solution.
\end{enumerate}

We note that a physical explanation of the observed breaks in the P-L relations is still an open question. 
\citet[][and references therein]{bhardwaj14} have provided possible theoretical interpretation of observed breaks in period-color and 
amplitude-color relations based on the interaction between hydrogen ionization front (HIF) and stellar photosphere (SP)
at different periods and phases. Moreover, the observed breaks may also be related to the changes in
the light curve structure as seen in the Fourier parameters at certain periods. For e.g., the Fourier amplitude
parameters for first-overtone mode Cepheids show a sharp dip at 2.5 days at optical bands \citep{bhardwaj14}. 
Similarly, the amplitude parameters show a distinct separation for wavelengths longer than $J$ as compared to 
optical bands around 20 days \citep{anupam15}. However, a correlation between observed breaks and the changes in
the Fourier parameters at various periods needs a further investigation. It will also be important to compare the
theoretical light curves generated from stellar pulsation models with observed light curve, quantitatively, to 
understand the physics behind these non-linearities.

These results will also have an importance from a stellar pulsation and evolution point of view as various 
non-linearities can be used to place constraints on the mass-luminosity (M-L) relations and instability strip 
topologies obeyed by Cepheids. Currently, theory suggests that the M-L relation is theoretically predicted to 
constraint the zero-point of the Period-Luminosity-Color relation for Classical Cepheids and in turn the P-L 
relation zero point \citep{bono99, bono00}. However, the LMC P-L relation has strong evidence of a break at 10 
days with statistically significantly different slopes on either side of 10 days. This is possible because for a 
fixed M-L relation obeyed by LMC Cepheids, the envelope structure changes with mass and hence with period 
\citep{kanbur95, smk5}. More specifically, the relative location of the interaction between the SP and HIF changes 
with period. This changing interaction leads to a physical mechanism that can cause changes in the slope of 
the P-L and/or P-C relations for a given M-L relation and fixed topology of the instability strip 
\citep{smk2, smk3, smk5}. The change in the topology of the instability 
strip with the metal abundance can also cause the difference in slopes of the LMC and SMC Cepheid P-L relations. It is 
also expected that the metallicity differences lead to different M-L relations obeyed by Cepheids in the Magellanic 
Clouds, which mainly affects the zero-point. However, each M-L pair has an envelope structure that changes with period
and hence a change in the SP/HIF interaction leading to possible changes in the slope of P-L relations. In this sense we feel 
that the M-L relations and the instability strip boundaries (or topology) are not completely independent of each other.

In a subsequent publication we plan to study P-L relations as a function of pulsation phase to further investigate the 
effect of non-linearity and possible metallicity dependence on the distance scale. However, we also note that
multiphase PL/PC relations for a given galaxy, say the LMC, show considerable changes in linearity, dispersion, 
slope and zero point \citep{smk1}. We also intend to investigate the physical mechanism for all these 
observed breaks in the P-L relations by a rigorous comparison with theoretical models. We believe that the 
theoretical stellar pulsation modeling of observed non-linearities, in principle, can provide 
important constraints on theories of stellar evolution and pulsation. In particular, these should be allied with comparison 
of the Fourier parameters of observed and theoretical Cepheid multiwavelength light curves \citep{anupam15}.

\section*{Acknowledgments}
\label{sec:ackno}
The authors acknowledge the Indo-U.S. Science and Technology Forum for the grant provided
to University of Delhi for the Joint Center for Analysis of Variable Star Data. We are very 
grateful to Adam Riess for his helpful guidance regarding the matrix formalism used to analyze SH0ES data. 
AB is thankful to the Council of Scientific and Industrial Research, New Delhi, for a Senior Research
Fellowship and acknowledges the support from Texas A\&M University, where some of this research was
carried out. LMM acknowledges support by the United States National Science Foundation through AST 
grant number 1211603 and by Texas A\&M University through a faculty start-up fund and the Mitchell-Heep-Munnerlyn
Endowed Career Enhancement Professorship in Physics or Astronomy. HPS thanks University of Delhi for a R\&D grant.
CCN thanks the funding from Ministry of Science and Technology (Taiwan) under the contract NSC101-2112-M-008-017-MY3 
and NSC104-2112-M-008-012-MY3. EEOI thanks Rafael S. de Souza and the Cosmostatistics Initiative (COIN) for 
insightful suggestions. EEOI is partially supported by the Brazilian agency CAPES (grant number 9229-13-2). 
This research was supported by the Munich Institute for Astro- and Particle Physics (MIAPP) of the DFG 
cluster of excellence ``Origin and Structure of the Universe''.

\bibliographystyle{mn2e}
\bibliography{cpapir_III}

\appendix
\section{Testimator Tables}
We provide the results of testimator analysis for fundamental and first-overtone mode Cepheids in the 
Tables~\ref{table:testimator_fu_1}, \ref{table:ss_testimator} and \ref{table:testimator_fo_1}.
The format of the tables and the meaning of each column header is similar to Table~\ref{table:simul_pl2}.
The number of independent and non-overlapping subsets are represented by $n$. Each subset consist of $N$
number of stars over a period range listed under the column $\log(P)$. The slope of P-L relation in each
subset is represented by $\hat{\beta}$. Also, $\beta_{0}$ and $\beta_{w}$ denote the initial testimator slope
and the updated testimator slope, respectively, after each hypothesis testing. The observed and critical value
of t-test statistics are denoted by $|t_{obs}|$ (estimated using equation~\ref{eq:tvalue}) and $t_{c}$ 
(obtained using the theoretical t-distribution for a confidence level of more than 95\%), respectively. 
Finally, $k$ is the ratio of $|t_{obs}|$ and $t_{c}$ and represents the probability of initial guess of 
the testimator being true. The value of $k<1$/$k>1$ leads to the decision of acceptance/rejection of the null
hypothesis (the slopes are equal in two subsets).

\begin{table*}
\begin{minipage}{1.0\hsize}
\begin{center}
\caption{Results of the testimator on P-L, P-W and P-C relations for FU mode Cepheids. The OGLE-III data 
are used in optical band relations.}
\label{table:testimator_fu_1}
\begin{tabular}{|l|c|c|c|c|c|c|c|c|c|c|}
\hline
\hline
Band &$n$  &  $\log(P)$&	$N$&	$\hat{\beta}$&	$\beta_{0}$&	$|t_{obs}|$&	$t_{c}$&	$k$&	Decision&	$\beta_{w}$\\
\hline
$V$&      1&   0.39900$-$0.45900   &         189&    -2.346$\pm$0.669     &       ---&       ---&       ---&       ---&       ---         &       ---\\
& 2&   0.45900$-$0.49200   &190&    -3.988$\pm$1.338     &    -2.346&     1.227&     2.765&     0.444&Accept~$H_{0}$&    -3.074\\
& 3&   0.49200$-$0.52900   &189&    -1.890$\pm$1.103     &    -3.074&     1.074&     2.765&     0.388&Accept~$H_{0}$&    -2.615\\
& 4&   0.52900$-$0.57500   &192&    -3.404$\pm$0.819     &    -2.615&     0.964&     2.765&     0.349&Accept~$H_{0}$&    -2.890\\
& 5&   0.57500$-$0.63600   &189&    -2.554$\pm$0.651     &    -2.890&     0.516&     2.765&     0.187&Accept~$H_{0}$&    -2.827\\
& 6&   0.63600$-$0.70800   &191&    -2.809$\pm$0.641     &    -2.827&     0.029&     2.765&     0.011&Accept~$H_{0}$&    -2.827\\
& 7&   0.70800$-$0.85600   &188&    -2.817$\pm$0.332     &    -2.827&     0.030&     2.766&     0.011&Accept~$H_{0}$&    -2.827\\
& 8&   0.85600$-$1.49200   &217&    -2.424$\pm$0.100     &    -2.827&     4.027&     2.761&     1.458&Reject~$H_{0}$&    -2.239\\
$I$&      1&   0.39900$-$0.45900   &         189&    -3.035$\pm$0.472     &       ---&       ---&       ---&       ---&       ---         &       ---\\
& 2&   0.45900$-$0.49200   &188&    -3.757$\pm$0.895     &    -3.035&     0.807&     2.766&     0.292&Accept~$H_{0}$&    -3.246\\
& 3&   0.49200$-$0.52900   &189&    -2.316$\pm$0.764     &    -3.246&     1.218&     2.765&     0.440&Accept~$H_{0}$&    -2.837\\
& 4&   0.52900$-$0.57600   &193&    -3.279$\pm$0.561     &    -2.837&     0.789&     2.765&     0.285&Accept~$H_{0}$&    -2.963\\
& 5&   0.57600$-$0.63600   &189&    -2.839$\pm$0.463     &    -2.963&     0.269&     2.765&     0.097&Accept~$H_{0}$&    -2.951\\
& 6&   0.63600$-$0.70800   &189&    -3.155$\pm$0.436     &    -2.951&     0.468&     2.765&     0.169&Accept~$H_{0}$&    -2.985\\
& 7&   0.70800$-$0.86000   &193&    -3.093$\pm$0.215     &    -2.985&     0.502&     2.765&     0.182&Accept~$H_{0}$&    -3.005\\
& 8&   0.86000$-$1.49200   &206&    -2.719$\pm$0.071     &    -3.005&     4.050&     2.763&     1.466&Reject~$H_{0}$&    -2.586\\
$J$&      1&   0.39900$-$0.48900   &         150&    -1.956$\pm$0.398     &       ---&       ---&       ---&       ---&       ---         &       ---\\
& 2&   0.48900$-$0.55500   &147&    -2.844$\pm$0.486     &    -1.956&     1.826&     2.610&     0.700&Accept~$H_{0}$&    -2.578\\
& 3&   0.55500$-$0.65900   &153&    -3.199$\pm$0.318     &    -2.578&     1.955&     2.608&     0.750&Accept~$H_{0}$&    -3.043\\
& 4&   0.65900$-$0.83500   &150&    -2.773$\pm$0.208     &    -3.043&     1.299&     2.609&     0.498&Accept~$H_{0}$&    -2.909\\
& 5&   0.83500$-$1.89700   &187&    -3.190$\pm$0.044     &    -2.909&     6.346&     2.602&     2.438&Reject~$H_{0}$&    -3.596\\
$H$&      1&   0.39900$-$0.48900   &         150&    -2.367$\pm$0.315     &       ---&       ---&       ---&       ---&       ---         &       ---\\
& 2&   0.48900$-$0.55500   &147&    -2.997$\pm$0.392     &    -2.367&     1.606&     2.610&     0.615&Accept~$H_{0}$&    -2.754\\
& 3&   0.55500$-$0.65900   &153&    -3.239$\pm$0.253     &    -2.754&     1.917&     2.608&     0.735&Accept~$H_{0}$&    -3.111\\
& 4&   0.65900$-$0.83500   &150&    -2.868$\pm$0.172     &    -3.111&     1.411&     2.609&     0.541&Accept~$H_{0}$&    -2.979\\
& 5&   0.83500$-$1.89700   &187&    -3.213$\pm$0.038     &    -2.979&     6.158&     2.602&     2.366&Reject~$H_{0}$&    -3.531\\
$K_{s}$&      1&   0.39900$-$0.48900   &         150&    -2.584$\pm$0.270     &       ---&       ---&       ---&       ---&       ---         &       ---\\
& 2&   0.48900$-$0.55500   &147&    -3.006$\pm$0.362     &    -2.584&     1.165&     2.610&     0.446&Accept~$H_{0}$&    -2.772\\
& 3&   0.55500$-$0.65900   &153&    -3.342$\pm$0.219     &    -2.772&     2.599&     2.608&     0.996&Accept~$H_{0}$&    -3.340\\
& 4&   0.65900$-$0.83500   &150&    -2.953$\pm$0.149     &    -3.340&     2.605&     2.609&     0.998&Accept~$H_{0}$&    -2.954\\
& 5&   0.83500$-$1.89700   &187&    -3.261$\pm$0.034     &    -2.954&     9.120&     2.602&     3.504&Reject~$H_{0}$&    -4.032\\
\hline
$W_{V,I}$&      1&   0.39900$-$0.45800   &         190&    -3.242$\pm$0.299     &       ---&       ---&       ---&       ---&       ---         &       ---\\
& 2&   0.45800$-$0.49000   &185&    -2.779$\pm$0.537     &    -3.242&     0.862&     2.766&     0.312&Accept~$H_{0}$&    -3.098\\
& 3&   0.49000$-$0.52600   &192&    -2.787$\pm$0.478     &    -3.098&     0.650&     2.765&     0.235&Accept~$H_{0}$&    -3.025\\
& 4&   0.52600$-$0.57000   &192&    -3.834$\pm$0.389     &    -3.025&     2.080&     2.765&     0.752&Accept~$H_{0}$&    -3.633\\
& 5&   0.57000$-$0.62800   &191&    -3.791$\pm$0.290     &    -3.633&     0.544&     2.765&     0.197&Accept~$H_{0}$&    -3.664\\
& 6&   0.62800$-$0.69500   &188&    -3.482$\pm$0.245     &    -3.664&     0.746&     2.766&     0.270&Accept~$H_{0}$&    -3.615\\
& 7&   0.69500$-$0.83500   &191&    -3.565$\pm$0.151     &    -3.615&     0.329&     2.765&     0.119&Accept~$H_{0}$&    -3.609\\
& 8&   0.83500$-$1.49200   &237&    -3.233$\pm$0.042     &    -3.609&     9.062&     2.759&     3.284&Reject~$H_{0}$&    -2.373\\
$W_{J,H}$&      1&   0.39900$-$0.48900   &         150&    -3.034$\pm$0.366     &       ---&       ---&       ---&       ---&       ---         &       ---\\
& 2&   0.48900$-$0.55500   &147&    -3.247$\pm$0.466     &    -3.034&     0.456&     2.610&     0.175&Accept~$H_{0}$&    -3.071\\
& 3&   0.55500$-$0.65900   &153&    -3.306$\pm$0.289     &    -3.071&     0.813&     2.608&     0.312&Accept~$H_{0}$&    -3.144\\
& 4&   0.65900$-$0.83500   &150&    -3.023$\pm$0.172     &    -3.144&     0.706&     2.609&     0.270&Accept~$H_{0}$&    -3.111\\
& 5&   0.83500$-$1.89700   &187&    -3.249$\pm$0.038     &    -3.111&     3.579&     2.602&     1.375&Reject~$H_{0}$&    -3.301\\
$W_{J,K_{s}}$&      1&   0.39900$-$0.48900   &         150&    -3.016$\pm$0.260     &       ---&       ---&       ---&       ---&       ---         &       ---\\
& 2&   0.48900$-$0.55500   &147&    -3.115$\pm$0.345     &    -3.016&     0.285&     2.610&     0.109&Accept~$H_{0}$&    -3.027\\
& 3&   0.55500$-$0.65900   &153&    -3.442$\pm$0.200     &    -3.027&     2.078&     2.608&     0.797&Accept~$H_{0}$&    -3.357\\
& 4&   0.65900$-$0.83500   &150&    -3.078$\pm$0.127     &    -3.357&     2.208&     2.609&     0.846&Accept~$H_{0}$&    -3.121\\
& 5&   0.83500$-$1.89700   &187&    -3.310$\pm$0.030     &    -3.121&     6.292&     2.602&     2.418&Reject~$H_{0}$&    -3.579\\
$W_{H,K_{s}}$&      1&   0.39900$-$0.48900   &         150&    -3.001$\pm$0.339     &       ---&       ---&       ---&       ---&       ---         &       ---\\
& 2&   0.48900$-$0.55500   &147&    -3.023$\pm$0.468     &    -3.001&     0.048&     2.610&     0.018&Accept~$H_{0}$&    -3.001\\
& 3&   0.55500$-$0.65900   &153&    -3.540$\pm$0.274     &    -3.001&     1.968&     2.608&     0.755&Accept~$H_{0}$&    -3.408\\
& 4&   0.65900$-$0.83500   &150&    -3.117$\pm$0.160     &    -3.408&     1.814&     2.609&     0.695&Accept~$H_{0}$&    -3.206\\
& 5&   0.83500$-$1.89700   &187&    -3.355$\pm$0.032     &    -3.206&     4.599&     2.602&     1.767&Reject~$H_{0}$&    -3.469\\
$W_{V,J}$&      1&   0.39900$-$0.49800   &         139&    -2.387$\pm$0.389     &       ---&       ---&       ---&       ---&       ---         &       ---\\
& 2&   0.49800$-$0.57000   &140&    -3.627$\pm$0.528     &    -2.387&     2.351&     2.611&     0.900&Accept~$H_{0}$&    -3.504\\
& 3&   0.57000$-$0.68100   &141&    -3.604$\pm$0.322     &    -3.504&     0.313&     2.611&     0.120&Accept~$H_{0}$&    -3.516\\
& 4&   0.68100$-$0.89300   &140&    -3.067$\pm$0.146     &    -3.516&     3.072&     2.611&     1.176&Reject~$H_{0}$&    -2.988\\
\hline
\end{tabular}
\end{center}
\end{minipage}
\end{table*}

\begin{table*}
\begin{minipage}{1.0\hsize}
\begin{center}
\begin{tabular}{|l|c|c|c|c|c|c|c|c|c|c|}
\multicolumn{11}{c}{{\bf Table~\ref{table:testimator_fu_1} :} Contd...}\\
\hline
\hline
Band &$n$  &  $\log(P)$&	$N$&	$\hat{\beta}$&	$\beta_{0}$&	$|t_{obs}|$&	$t_{c}$&	$k$&	Decision&	$\beta_{w}$\\
\hline
$W_{V,H}$&      1&   0.39900$-$0.49800   &         140&    -2.385$\pm$0.334     &       ---&       ---&       ---&       ---&       ---         &       ---\\
& 2&   0.49800$-$0.57000   &140&    -3.456$\pm$0.440     &    -2.385&     2.434&     2.611&     0.932&Accept~$H_{0}$&    -3.383\\
& 3&   0.57000$-$0.68000   &140&    -3.685$\pm$0.283     &    -3.383&     1.066&     2.611&     0.408&Accept~$H_{0}$&    -3.506\\
& 4&   0.68000$-$0.89000   &140&    -3.064$\pm$0.135     &    -3.506&     3.266&     2.611&     1.251&Reject~$H_{0}$&    -2.954\\
$W_{V,K_{s}}$&      1&   0.39900$-$0.49800   &         139&    -2.972$\pm$0.227     &       ---&       ---&       ---&       ---&       ---         &       ---\\
& 2&   0.49800$-$0.57000   &140&    -3.323$\pm$0.335     &    -2.972&     1.047&     2.611&     0.401&Accept~$H_{0}$&    -3.113\\
& 3&   0.57000$-$0.68100   &140&    -3.485$\pm$0.204     &    -3.113&     1.827&     2.611&     0.700&Accept~$H_{0}$&    -3.373\\
& 4&   0.68100$-$0.89500   &141&    -3.162$\pm$0.099     &    -3.373&     2.129&     2.611&     0.815&Accept~$H_{0}$&    -3.201\\
& 5&   0.89500$-$1.89700   &147&    -3.333$\pm$0.034     &    -3.201&     3.938&     2.610&     1.509&Reject~$H_{0}$&    -3.400\\
$W_{I,J}$&      1&   0.39900$-$0.49800   &         140&    -2.354$\pm$0.508     &       ---&       ---&       ---&       ---&       ---         &       ---\\
& 2&   0.49800$-$0.57000   &139&    -3.835$\pm$0.660     &    -2.354&     2.242&     2.612&     0.859&Accept~$H_{0}$&    -3.625\\
& 3&   0.57000$-$0.68000   &140&    -3.694$\pm$0.414     &    -3.625&     0.167&     2.611&     0.064&Accept~$H_{0}$&    -3.630\\
& 4&   0.68000$-$0.89000   &140&    -2.992$\pm$0.182     &    -3.630&     3.498&     2.611&     1.340&Reject~$H_{0}$&    -2.775\\
$W_{I,H}$&      1&   0.39900$-$0.49800   &         140&    -2.643$\pm$0.313     &       ---&       ---&       ---&       ---&       ---         &       ---\\
& 2&   0.49800$-$0.57000   &140&    -3.685$\pm$0.418     &    -2.643&     2.494&     2.611&     0.955&Accept~$H_{0}$&    -3.638\\
& 3&   0.57000$-$0.68000   &139&    -3.329$\pm$0.267     &    -3.638&     1.158&     2.612&     0.443&Accept~$H_{0}$&    -3.501\\
& 4&   0.68000$-$0.89000   &140&    -2.996$\pm$0.127     &    -3.501&     3.959&     2.611&     1.516&Reject~$H_{0}$&    -2.736\\
$W_{I,K_{s}}$&      1&   0.39900$-$0.49800   &         139&    -2.976$\pm$0.241     &       ---&       ---&       ---&       ---&       ---         &       ---\\
& 2&   0.49800$-$0.57000   &140&    -3.251$\pm$0.362     &    -2.976&     0.759&     2.611&     0.291&Accept~$H_{0}$&    -3.056\\
& 3&   0.57000$-$0.68100   &140&    -3.364$\pm$0.207     &    -3.056&     1.485&     2.611&     0.569&Accept~$H_{0}$&    -3.231\\
& 4&   0.68100$-$0.89500   &141&    -3.147$\pm$0.103     &    -3.231&     0.815&     2.611&     0.312&Accept~$H_{0}$&    -3.205\\
& 5&   0.89500$-$1.89700   &148&    -3.346$\pm$0.034     &    -3.205&     4.112&     2.609&     1.576&Reject~$H_{0}$&    -3.428\\
$W^H_{V,I}$&      1&   0.39900$-$0.49800   &         139&    -2.808$\pm$0.249     &       ---&       ---&       ---&       ---&       ---         &       ---\\
& 2&   0.49800$-$0.57000   &141&    -3.377$\pm$0.338     &    -2.808&     1.683&     2.611&     0.644&Accept~$H_{0}$&    -3.175\\
& 3&   0.57000$-$0.68100   &140&    -3.373$\pm$0.209     &    -3.175&     0.944&     2.611&     0.362&Accept~$H_{0}$&    -3.246\\
& 4&   0.68100$-$0.89300   &140&    -3.094$\pm$0.108     &    -3.246&     1.411&     2.611&     0.540&Accept~$H_{0}$&    -3.164\\
& 5&   0.89300$-$1.89700   &148&    -3.313$\pm$0.034     &    -3.164&     4.347&     2.609&     1.666&Reject~$H_{0}$&    -3.412\\
\hline
$V-I$&      1&   0.39900$-$0.45900   &         190&     0.316$\pm$0.239     &       ---&       ---&       ---&       ---&       ---         &       ---\\
& 2&   0.45900$-$0.49100   &182&    -0.210$\pm$0.518     &     0.316&     1.013&     2.767&     0.366&Accept~$H_{0}$&     0.123\\
& 3&   0.49100$-$0.52800   &197&     0.244$\pm$0.403     &     0.123&     0.301&     2.764&     0.109&Accept~$H_{0}$&     0.136\\
& 4&   0.52800$-$0.57400   &190&     0.232$\pm$0.320     &     0.136&     0.298&     2.765&     0.108&Accept~$H_{0}$&     0.147\\
& 5&   0.57400$-$0.63300   &187&     0.314$\pm$0.269     &     0.147&     0.623&     2.766&     0.225&Accept~$H_{0}$&     0.184\\
& 6&   0.63300$-$0.70100   &191&     0.445$\pm$0.242     &     0.184&     1.075&     2.765&     0.389&Accept~$H_{0}$&     0.286\\
& 7&   0.70100$-$0.84600   &193&     0.361$\pm$0.111     &     0.286&     0.677&     2.765&     0.245&Accept~$H_{0}$&     0.304\\
& 8&   0.84600$-$1.49200   &226&     0.350$\pm$0.035     &     0.304&     1.325&     2.760&     0.480&Accept~$H_{0}$&     0.326\\
$J-H$&      1&   0.39900$-$0.49200   &         130&     0.249$\pm$0.148     &       ---&       ---&       ---&       ---&       ---         &       ---\\
& 2&   0.49200$-$0.56100   &130&     0.024$\pm$0.202     &     0.249&     1.121&     2.614&     0.429&Accept~$H_{0}$&     0.153\\
& 3&   0.56100$-$0.66600   &130&    -0.087$\pm$0.119     &     0.153&     2.013&     2.614&     0.770&Accept~$H_{0}$&    -0.032\\
& 4&   0.66600$-$0.83800   &129&     0.019$\pm$0.066     &    -0.032&     0.776&     2.614&     0.297&Accept~$H_{0}$&    -0.017\\
& 5&   0.83800$-$1.89700   &178&     0.013$\pm$0.014     &    -0.017&     2.142&     2.604&     0.823&Accept~$H_{0}$&     0.008\\
$J-K_{s}$&      1&   0.39900$-$0.49400   &         130&     0.368$\pm$0.164     &       ---&       ---&       ---&       ---&       ---         &       ---\\
& 2&   0.49400$-$0.56200   &130&    -0.213$\pm$0.216     &     0.368&     2.689&     2.614&     1.029&Reject~$H_{0}$&    -0.230\\
$H-K_{s}$&      1&   0.39900$-$0.49200   &         130&     0.221$\pm$0.108     &       ---&       ---&       ---&       ---&       ---         &       ---\\
& 2&   0.49200$-$0.56100   &130&    -0.034$\pm$0.138     &     0.221&     1.847&     2.614&     0.707&Accept~$H_{0}$&     0.041\\
& 3&   0.56100$-$0.66500   &128&     0.047$\pm$0.101     &     0.041&     0.066&     2.615&     0.025&Accept~$H_{0}$&     0.041\\
& 4&   0.66500$-$0.83800   &131&     0.100$\pm$0.048     &     0.041&     1.215&     2.614&     0.465&Accept~$H_{0}$&     0.068\\
& 5&   0.83800$-$1.89700   &177&     0.056$\pm$0.008     &     0.068&     1.461&     2.604&     0.561&Accept~$H_{0}$&     0.061\\
$V-J$&      1&   0.39900$-$0.49100   &         127&     0.780$\pm$0.398     &       ---&       ---&       ---&       ---&       ---         &       ---\\
& 2&   0.49100$-$0.55800   &132&     0.868$\pm$0.552     &     0.780&     0.159&     2.614&     0.061&Accept~$H_{0}$&     0.786\\
& 3&   0.55800$-$0.66500   &131&     0.782$\pm$0.371     &     0.786&     0.011&     2.614&     0.004&Accept~$H_{0}$&     0.786\\
& 4&   0.66500$-$0.83800   &129&     0.432$\pm$0.231     &     0.786&     1.529&     2.614&     0.585&Accept~$H_{0}$&     0.579\\
& 5&   0.83800$-$1.89700   &178&     0.391$\pm$0.048     &     0.579&     3.892&     2.604&     1.495&Reject~$H_{0}$&     0.298\\
$V-H$&      1&   0.39900$-$0.49100   &         126&     1.034$\pm$0.445     &       ---&       ---&       ---&       ---&       ---         &       ---\\
& 2&   0.49100$-$0.55900   &134&     1.158$\pm$0.612     &     1.034&     0.202&     2.613&     0.077&Accept~$H_{0}$&     1.044\\
& 3&   0.55900$-$0.66500   &128&     0.983$\pm$0.412     &     1.044&     0.148&     2.615&     0.057&Accept~$H_{0}$&     1.041\\
& 4&   0.66500$-$0.84100   &132&     0.377$\pm$0.249     &     1.041&     2.665&     2.614&     1.020&Reject~$H_{0}$&     0.364\\
$V-K_{s}$&      1&   0.39900$-$0.49200   &         130&     1.030$\pm$0.409     &       ---&       ---&       ---&       ---&       ---         &       ---\\
& 2&   0.49200$-$0.55900   &129&     0.729$\pm$0.672     &     1.030&     0.448&     2.614&     0.171&Accept~$H_{0}$&     0.978\\
& 3&   0.55900$-$0.66600   &131&     1.117$\pm$0.403     &     0.978&     0.345&     2.614&     0.132&Accept~$H_{0}$&     0.997\\
& 4&   0.66600$-$0.84700   &130&     0.371$\pm$0.241     &     0.997&     2.595&     2.614&     0.993&Accept~$H_{0}$&     0.375\\
& 5&   0.84700$-$1.89700   &173&     0.446$\pm$0.062     &     0.375&     1.144&     2.605&     0.439&Accept~$H_{0}$&     0.406\\
\hline
\end{tabular}
\end{center}
\end{minipage}
\end{table*}

\begin{table*}
\begin{minipage}{1.0\hsize}
\begin{center}
\begin{tabular}{|l|c|c|c|c|c|c|c|c|c|c|}
\multicolumn{11}{c}{{\bf Table~\ref{table:testimator_fu_1} :} Contd...}\\
\hline
\hline
Band &$n$  &  $\log(P)$&	$N$&	$\hat{\beta}$&	$\beta_{0}$&	$|t_{obs}|$&	$t_{c}$&	$k$&	Decision&	$\beta_{w}$\\
\hline

$I-J$&      1&   0.39900$-$0.49800   &         139&     0.075$\pm$0.232     &       ---&       ---&       ---&       ---&       ---         &       ---\\
& 2&   0.49800$-$0.57000   &138&     0.374$\pm$0.295     &     0.075&     1.016&     2.612&     0.389&Accept~$H_{0}$&     0.191\\
& 3&   0.57000$-$0.68200   &142&     0.261$\pm$0.199     &     0.191&     0.350&     2.611&     0.134&Accept~$H_{0}$&     0.201\\
& 4&   0.68200$-$0.91300   &141&     0.076$\pm$0.096     &     0.201&     1.301&     2.611&     0.498&Accept~$H_{0}$&     0.138\\
& 5&   0.91300$-$1.89700   &141&     0.152$\pm$0.032     &     0.138&     0.429&     2.611&     0.164&Accept~$H_{0}$&     0.141\\
$I-H$&      1&   0.39900$-$0.49800   &         139&     0.353$\pm$0.277     &       ---&       ---&       ---&       ---&       ---         &       ---\\
& 2&   0.49800$-$0.57000   &138&     0.609$\pm$0.355     &     0.353&     0.720&     2.612&     0.276&Accept~$H_{0}$&     0.423\\
& 3&   0.57000$-$0.68300   &143&     0.074$\pm$0.223     &     0.423&     1.563&     2.611&     0.599&Accept~$H_{0}$&     0.214\\
& 4&   0.68300$-$0.91400   &140&     0.015$\pm$0.115     &     0.214&     1.732&     2.611&     0.663&Accept~$H_{0}$&     0.082\\
& 5&   0.91400$-$1.89700   &140&     0.156$\pm$0.040     &     0.082&     1.850&     2.611&     0.708&Accept~$H_{0}$&     0.134\\
$I-K_{s}$&      1&   0.39900$-$0.49100   &         126&     0.642$\pm$0.294     &       ---&       ---&       ---&       ---&       ---         &       ---\\
& 2&   0.49100$-$0.55900   &133&     0.551$\pm$0.360     &     0.642&     0.253&     2.613&     0.097&Accept~$H_{0}$&     0.633\\
& 3&   0.55900$-$0.66500   &129&     0.524$\pm$0.255     &     0.633&     0.429&     2.614&     0.164&Accept~$H_{0}$&     0.615\\
& 4&   0.66500$-$0.84000   &132&     0.139$\pm$0.157     &     0.615&     3.026&     2.614&     1.158&Reject~$H_{0}$&     0.064\\
\hline
\end{tabular}
\end{center}
\end{minipage}
\end{table*}

\begin{table*}
\begin{minipage}{1.0\hsize}
\begin{center}
\caption{Results of the testimator on P-L, P-W and P-C relations for FU mode Cepheids using the OGLE-III-SS data.}
\label{table:ss_testimator}
\begin{tabular}{|l|c|c|c|c|c|c|c|c|c|c|}
\hline
\hline
Band &$n$  &  $\log(P)$&	$N$&	$\hat{\beta}$&	$\beta_{0}$&	$|t_{obs}|$&	$t_{c}$&	$k$&	Decision&$\beta_{w}$\\
\hline
$V$&      1&   0.39900$-$0.45900   &         189&    -2.346$\pm$0.669     &       ---&       ---&       ---&       ---&       ---         &       ---\\
& 2&   0.45900$-$0.49200   &190&    -3.988$\pm$1.338     &    -2.346&     1.227&     2.765&     0.444&Accept~$H_{0}$&    -3.074\\
& 3&   0.49200$-$0.52900   &189&    -1.890$\pm$1.103     &    -3.074&     1.074&     2.765&     0.388&Accept~$H_{0}$&    -2.615\\
& 4&   0.52900$-$0.57400   &189&    -2.969$\pm$0.878     &    -2.615&     0.403&     2.765&     0.146&Accept~$H_{0}$&    -2.666\\
& 5&   0.57400$-$0.63500   &193&    -2.554$\pm$0.666     &    -2.666&     0.168&     2.765&     0.061&Accept~$H_{0}$&    -2.659\\
& 6&   0.63500$-$0.70500   &188&    -2.606$\pm$0.656     &    -2.659&     0.081&     2.766&     0.029&Accept~$H_{0}$&    -2.658\\
& 7&   0.70500$-$0.85300   &192&    -2.750$\pm$0.328     &    -2.658&     0.281&     2.765&     0.102&Accept~$H_{0}$&    -2.667\\
& 8&   0.85300$-$1.71900   &238&    -2.680$\pm$0.074     &    -2.667&     0.168&     2.759&     0.061&Accept~$H_{0}$&    -2.668\\
$I$&      1&   0.39900$-$0.45900   &         189&    -3.035$\pm$0.472     &       ---&       ---&       ---&       ---&       ---         &       ---\\
& 2&   0.45900$-$0.49200   &189&    -3.903$\pm$0.911     &    -3.035&     0.953&     2.765&     0.345&Accept~$H_{0}$&    -3.334\\
& 3&   0.49200$-$0.52900   &189&    -2.316$\pm$0.764     &    -3.334&     1.333&     2.765&     0.482&Accept~$H_{0}$&    -2.843\\
& 4&   0.52900$-$0.57600   &193&    -3.279$\pm$0.561     &    -2.843&     0.776&     2.765&     0.281&Accept~$H_{0}$&    -2.966\\
& 5&   0.57600$-$0.63600   &189&    -2.839$\pm$0.463     &    -2.966&     0.275&     2.765&     0.099&Accept~$H_{0}$&    -2.953\\
& 6&   0.63600$-$0.70800   &190&    -3.301$\pm$0.441     &    -2.953&     0.789&     2.765&     0.285&Accept~$H_{0}$&    -3.052\\
& 7&   0.70800$-$0.85900   &190&    -3.115$\pm$0.226     &    -3.052&     0.277&     2.765&     0.100&Accept~$H_{0}$&    -3.059\\
& 8&   0.85900$-$1.71900   &230&    -2.949$\pm$0.052     &    -3.059&     2.124&     2.760&     0.769&Accept~$H_{0}$&    -2.974\\
$W_{V,I}$&      1&   0.39900$-$0.45800   &         190&    -3.242$\pm$0.299     &       ---&       ---&       ---&       ---&       ---         &       ---\\
& 2&   0.45800$-$0.49000   &185&    -2.779$\pm$0.537     &    -3.242&     0.862&     2.766&     0.312&Accept~$H_{0}$&    -3.098\\
& 3&   0.49000$-$0.52600   &193&    -2.767$\pm$0.489     &    -3.098&     0.675&     2.765&     0.244&Accept~$H_{0}$&    -3.017\\
& 4&   0.52600$-$0.57000   &192&    -3.834$\pm$0.389     &    -3.017&     2.099&     2.765&     0.759&Accept~$H_{0}$&    -3.637\\
& 5&   0.57000$-$0.62700   &188&    -3.812$\pm$0.299     &    -3.637&     0.584&     2.766&     0.211&Accept~$H_{0}$&    -3.674\\
& 6&   0.62700$-$0.69500   &191&    -3.462$\pm$0.238     &    -3.674&     0.893&     2.765&     0.323&Accept~$H_{0}$&    -3.605\\
& 7&   0.69500$-$0.83500   &191&    -3.565$\pm$0.151     &    -3.605&     0.265&     2.765&     0.096&Accept~$H_{0}$&    -3.602\\
& 8&   0.83500$-$1.72400   &259&    -3.326$\pm$0.029     &    -3.602&     9.574&     2.757&     3.473&Reject~$H_{0}$&    -2.645\\
$V-I$&      1&   0.39900$-$0.45900   &         190&     0.316$\pm$0.239     &       ---&       ---&       ---&       ---&       ---         &       ---\\
& 2&   0.45900$-$0.49100   &182&    -0.210$\pm$0.518     &     0.316&     1.013&     2.767&     0.366&Accept~$H_{0}$&     0.123\\
& 3&   0.49100$-$0.52800   &197&     0.244$\pm$0.403     &     0.123&     0.301&     2.764&     0.109&Accept~$H_{0}$&     0.136\\
& 4&   0.52800$-$0.57400   &190&     0.232$\pm$0.320     &     0.136&     0.298&     2.765&     0.108&Accept~$H_{0}$&     0.147\\
& 5&   0.57400$-$0.63300   &187&     0.314$\pm$0.269     &     0.147&     0.623&     2.766&     0.225&Accept~$H_{0}$&     0.184\\
& 6&   0.63300$-$0.70100   &191&     0.445$\pm$0.242     &     0.184&     1.075&     2.765&     0.389&Accept~$H_{0}$&     0.286\\
& 7&   0.70100$-$0.84600   &193&     0.361$\pm$0.111     &     0.286&     0.677&     2.765&     0.245&Accept~$H_{0}$&     0.304\\
& 8&   0.84600$-$1.72400   &249&     0.252$\pm$0.025     &     0.304&     2.046&     2.758&     0.742&Accept~$H_{0}$&     0.266\\
\hline
\end{tabular}
\end{center}
\end{minipage}
\end{table*}

\begin{table*}
\begin{minipage}{1.0\hsize}
\begin{center}
\caption{Results of the testimator on P-L, P-W and P-C relations for FO mode Cepheids. The OGLE-III data 
are used in optical band relations.}
\label{table:testimator_fo_1}
\begin{tabular}{|l|c|c|c|c|c|c|c|c|c|c|}
\hline
\hline
Band &$n$  &  $\log(P)$&	$N$&	$\hat{\beta}$&	$\beta_{0}$&	$|t_{obs}|$&	$t_{c}$&	$k$&	Decision&	$\beta_{w}$\\
\hline
\hline
$V$&      1&  -0.15100$-$0.18100   &         210&    -3.194$\pm$0.118     &       ---&       ---&       ---&       ---&       ---         &       ---\\
& 2&   0.18100$-$0.28500   &208&    -2.931$\pm$0.447     &    -3.194&     0.588&     2.600&     0.226&Accept~$H_{0}$&    -3.134\\
& 3&   0.28500$-$0.33700   &211&    -3.897$\pm$0.895     &    -3.134&     0.852&     2.599&     0.328&Accept~$H_{0}$&    -3.384\\
& 4&   0.33700$-$0.43400   &211&    -4.285$\pm$0.404     &    -3.384&     2.232&     2.599&     0.859&Accept~$H_{0}$&    -4.158\\
& 5&   0.43400$-$0.77200   &232&    -2.721$\pm$0.136     &    -4.158&    10.606&     2.597&     4.084&Reject~$H_{0}$&     1.711\\
$I$&      1&  -0.15100$-$0.18100   &         209&    -3.238$\pm$0.087     &       ---&       ---&       ---&       ---&       ---         &       ---\\
& 2&   0.18100$-$0.28600   &210&    -3.171$\pm$0.310     &    -3.238&     0.216&     2.599&     0.083&Accept~$H_{0}$&    -3.233\\
& 3&   0.28600$-$0.33700   &208&    -3.694$\pm$0.617     &    -3.233&     0.747&     2.600&     0.288&Accept~$H_{0}$&    -3.365\\
& 4&   0.33700$-$0.43400   &213&    -4.013$\pm$0.298     &    -3.365&     2.172&     2.599&     0.836&Accept~$H_{0}$&    -3.906\\
& 5&   0.43400$-$0.77200   &231&    -2.932$\pm$0.097     &    -3.906&    10.012&     2.597&     3.855&Reject~$H_{0}$&    -0.152\\
$J$&      1&  -0.14100$-$0.25700   &         110&    -3.435$\pm$0.128     &       ---&       ---&       ---&       ---&       ---         &       ---\\
& 2&   0.25700$-$0.32300   &110&    -2.335$\pm$0.938     &    -3.435&     1.173&     2.539&     0.462&Accept~$H_{0}$&    -2.927\\
& 3&   0.32300$-$0.38600   &110&    -4.221$\pm$0.717     &    -2.927&     1.806&     2.539&     0.711&Accept~$H_{0}$&    -3.847\\
& 4&   0.38600$-$0.77200   &144&    -3.040$\pm$0.093     &    -3.847&     8.689&     2.529&     3.435&Reject~$H_{0}$&    -1.073\\
$H$&      1&  -0.14100$-$0.25700   &         110&    -3.405$\pm$0.105     &       ---&       ---&       ---&       ---&       ---         &       ---\\
& 2&   0.25700$-$0.32300   &110&    -2.716$\pm$0.658     &    -3.405&     1.046&     2.539&     0.412&Accept~$H_{0}$&    -3.121\\
& 3&   0.32300$-$0.38600   &110&    -4.169$\pm$0.539     &    -3.121&     1.944&     2.539&     0.766&Accept~$H_{0}$&    -3.924\\
& 4&   0.38600$-$0.77200   &144&    -3.068$\pm$0.067     &    -3.924&    12.712&     2.529&     5.026&Reject~$H_{0}$&     0.378\\
$K_{s}$&      1&  -0.14100$-$0.25700   &         110&    -3.209$\pm$0.094     &       ---&       ---&       ---&       ---&       ---         &       ---\\
& 2&   0.25700$-$0.32300   &110&    -3.054$\pm$0.560     &    -3.209&     0.277&     2.539&     0.109&Accept~$H_{0}$&    -3.192\\
& 3&   0.32300$-$0.38600   &110&    -3.696$\pm$0.454     &    -3.192&     1.110&     2.539&     0.437&Accept~$H_{0}$&    -3.413\\
& 4&   0.38600$-$0.77200   &144&    -3.138$\pm$0.057     &    -3.413&     4.807&     2.529&     1.900&Reject~$H_{0}$&    -2.890\\
\hline
$W_{V,I}$&      1&  -0.15100$-$0.19400   &         210&    -3.429$\pm$0.050     &       ---&       ---&       ---&       ---&       ---         &       ---\\
& 2&   0.19400$-$0.28800   &206&    -3.754$\pm$0.181     &    -3.429&     1.793&     2.600&     0.690&Accept~$H_{0}$&    -3.653\\
& 3&   0.28800$-$0.34000   &209&    -3.950$\pm$0.321     &    -3.653&     0.925&     2.600&     0.356&Accept~$H_{0}$&    -3.759\\
& 4&   0.34000$-$0.44100   &215&    -3.503$\pm$0.134     &    -3.759&     1.916&     2.599&     0.737&Accept~$H_{0}$&    -3.570\\
& 5&   0.44100$-$0.77200   &223&    -3.388$\pm$0.059     &    -3.570&     3.101&     2.598&     1.194&Reject~$H_{0}$&    -3.352\\
$W_{J,H}$&      1&  -0.14100$-$0.25700   &         110&    -3.355$\pm$0.146     &       ---&       ---&       ---&       ---&       ---         &       ---\\
& 2&   0.25700$-$0.32300   &110&    -3.334$\pm$0.657     &    -3.355&     0.032&     2.539&     0.013&Accept~$H_{0}$&    -3.355\\
& 3&   0.32300$-$0.38600   &110&    -4.084$\pm$0.666     &    -3.355&     1.095&     2.539&     0.431&Accept~$H_{0}$&    -3.669\\
& 4&   0.38600$-$0.77200   &144&    -3.113$\pm$0.086     &    -3.669&     6.438&     2.529&     2.545&Reject~$H_{0}$&    -2.253\\
$W_{J,K_{s}}$&      1&  -0.14100$-$0.25700   &         110&    -3.054$\pm$0.097     &       ---&       ---&       ---&       ---&       ---         &       ---\\
& 2&   0.25700$-$0.32300   &110&    -3.552$\pm$0.480     &    -3.054&     1.039&     2.539&     0.409&Accept~$H_{0}$&    -3.257\\
& 3&   0.32300$-$0.38600   &110&    -3.335$\pm$0.474     &    -3.257&     0.163&     2.539&     0.064&Accept~$H_{0}$&    -3.262\\
& 4&   0.38600$-$0.77200   &144&    -3.205$\pm$0.056     &    -3.262&     1.019&     2.529&     0.403&Accept~$H_{0}$&    -3.239\\
$W_{H,K_{s}}$&      1&  -0.14100$-$0.25700   &         110&    -2.834$\pm$0.138     &       ---&       ---&       ---&       ---&       ---         &       ---\\
& 2&   0.25700$-$0.32300   &110&    -3.700$\pm$0.683     &    -2.834&     1.268&     2.539&     0.499&Accept~$H_{0}$&    -3.267\\
& 3&   0.32300$-$0.38600   &110&    -2.785$\pm$0.637     &    -3.267&     0.756&     2.539&     0.298&Accept~$H_{0}$&    -3.123\\
& 4&   0.38600$-$0.77200   &144&    -3.272$\pm$0.082     &    -3.123&     1.817&     2.529&     0.718&Accept~$H_{0}$&    -3.230\\
$W_{V,J}$&      1&  -0.14100$-$0.24900   &         100&    -3.483$\pm$0.133     &       ---&       ---&       ---&       ---&       ---         &       ---\\
& 2&   0.24900$-$0.32700   & 99&    -3.947$\pm$0.711     &    -3.483&     0.652&     2.544&     0.256&Accept~$H_{0}$&    -3.602\\
& 3&   0.32700$-$0.39500   &100&    -3.548$\pm$0.665     &    -3.602&     0.082&     2.544&     0.032&Accept~$H_{0}$&    -3.600\\
& 4&   0.39500$-$0.77200   &125&    -3.106$\pm$0.111     &    -3.600&     4.461&     2.534&     1.760&Reject~$H_{0}$&    -2.730\\
$W_{V,H}$&      1&  -0.14100$-$0.24900   &         100&    -3.441$\pm$0.116     &       ---&       ---&       ---&       ---&       ---         &       ---\\
& 2&   0.24900$-$0.32800   &100&    -3.365$\pm$0.555     &    -3.441&     0.137&     2.544&     0.054&Accept~$H_{0}$&    -3.437\\
& 3&   0.32800$-$0.39800   &100&    -3.474$\pm$0.492     &    -3.437&     0.075&     2.544&     0.029&Accept~$H_{0}$&    -3.438\\
& 4&   0.39800$-$0.77200   &123&    -3.058$\pm$0.090     &    -3.438&     4.227&     2.535&     1.668&Reject~$H_{0}$&    -2.805\\
$W_{V,K_{s}}$&      1&  -0.14100$-$0.25000   &         100&    -3.191$\pm$0.089     &       ---&       ---&       ---&       ---&       ---         &       ---\\
& 2&   0.25000$-$0.32900   &100&    -3.230$\pm$0.410     &    -3.191&     0.096&     2.544&     0.038&Accept~$H_{0}$&    -3.193\\
& 3&   0.32900$-$0.39900   &100&    -3.017$\pm$0.402     &    -3.193&     0.438&     2.544&     0.172&Accept~$H_{0}$&    -3.162\\
& 4&   0.39900$-$0.77200   &122&    -3.152$\pm$0.060     &    -3.162&     0.178&     2.535&     0.070&Accept~$H_{0}$&    -3.162\\
$W_{I,J}$&      1&  -0.14100$-$0.24900   &         100&    -3.454$\pm$0.176     &       ---&       ---&       ---&       ---&       ---         &       ---\\
& 2&   0.24900$-$0.32800   &100&    -3.724$\pm$0.858     &    -3.454&     0.315&     2.544&     0.124&Accept~$H_{0}$&    -3.487\\
& 3&   0.32800$-$0.39800   &100&    -2.896$\pm$0.847     &    -3.487&     0.697&     2.544&     0.274&Accept~$H_{0}$&    -3.325\\
& 4&   0.39800$-$0.77200   &123&    -3.010$\pm$0.141     &    -3.325&     2.238&     2.535&     0.883&Accept~$H_{0}$&    -3.047\\
$W_{I,H}$&      1&  -0.14100$-$0.25000   &         100&    -3.377$\pm$0.107     &       ---&       ---&       ---&       ---&       ---         &       ---\\
& 2&   0.25000$-$0.32900   &100&    -2.921$\pm$0.500     &    -3.377&     0.911&     2.544&     0.358&Accept~$H_{0}$&    -3.214\\
& 3&   0.32900$-$0.39800   &100&    -3.082$\pm$0.497     &    -3.214&     0.264&     2.544&     0.104&Accept~$H_{0}$&    -3.200\\
& 4&   0.39800$-$0.77200   &122&    -3.090$\pm$0.071     &    -3.200&     1.554&     2.535&     0.613&Accept~$H_{0}$&    -3.133\\
\hline
\end{tabular}
\end{center}
\end{minipage}
\end{table*}
\begin{table*}
\begin{minipage}{1.0\hsize}
\begin{center}
\begin{tabular}{|l|c|c|c|c|c|c|c|c|c|c|}
\multicolumn{11}{c}{{\bf Table~\ref{table:testimator_fo_1} :} Contd...}\\
\hline
\hline
Band &$n$  &  $\log(P)$&	$N$&	$\hat{\beta}$&	$\beta_{0}$&	$|t_{obs}|$&	$t_{c}$&	$k$&	Decision&	$\beta_{w}$\\
\hline
$W_{I,K_{s}}$&      1&  -0.14100$-$0.25000   &         100&    -3.153$\pm$0.094     &       ---&       ---&       ---&       ---&       ---         &       ---\\
& 2&   0.25000$-$0.33000   & 99&    -3.150$\pm$0.396     &    -3.153&     0.008&     2.544&     0.003&Accept~$H_{0}$&    -3.153\\
& 3&   0.33000$-$0.40300   &101&    -2.965$\pm$0.393     &    -3.153&     0.480&     2.543&     0.189&Accept~$H_{0}$&    -3.118\\
& 4&   0.40300$-$0.77200   &120&    -3.145$\pm$0.064     &    -3.118&     0.425&     2.536&     0.168&Accept~$H_{0}$&    -3.122\\
$W^H_{V,I}$&      1&  -0.14100$-$0.25000   &         100&    -3.429$\pm$0.086     &       ---&       ---&       ---&       ---&       ---         &       ---\\
& 2&   0.25000$-$0.32800   &100&    -3.141$\pm$0.414     &    -3.429&     0.695&     2.544&     0.273&Accept~$H_{0}$&    -3.350\\
& 3&   0.32800$-$0.39500   & 99&    -3.523$\pm$0.422     &    -3.350&     0.409&     2.544&     0.161&Accept~$H_{0}$&    -3.378\\
& 4&   0.39500$-$0.77200   &124&    -3.148$\pm$0.058     &    -3.378&     3.968&     2.535&     1.566&Reject~$H_{0}$&    -3.019\\
\hline
$V-I$&      1&  -0.15100$-$0.18100   &         210&     0.052$\pm$0.042     &       ---&       ---&       ---&       ---&       ---         &       ---\\
& 2&   0.18100$-$0.28500   &209&     0.242$\pm$0.152     &     0.052&     1.244&     2.600&     0.478&Accept~$H_{0}$&     0.143\\
& 3&   0.28500$-$0.33800   &211&    -0.112$\pm$0.311     &     0.143&     0.821&     2.599&     0.316&Accept~$H_{0}$&     0.063\\
& 4&   0.33800$-$0.43300   &210&    -0.158$\pm$0.157     &     0.063&     1.406&     2.599&     0.541&Accept~$H_{0}$&    -0.057\\
& 5&   0.43300$-$0.77200   &234&     0.273$\pm$0.047     &    -0.057&     7.007&     2.597&     2.698&Reject~$H_{0}$&     0.832\\
$J-H$&      1&  -0.14100$-$0.25700   &         100&    -0.089$\pm$0.061     &       ---&       ---&       ---&       ---&       ---         &       ---\\
& 2&   0.25700$-$0.33200   & 99&     0.038$\pm$0.248     &    -0.089&     0.514&     2.544&     0.202&Accept~$H_{0}$&    -0.063\\
& 3&   0.33200$-$0.40800   &101&    -0.103$\pm$0.229     &    -0.063&     0.171&     2.543&     0.067&Accept~$H_{0}$&    -0.066\\
& 4&   0.40800$-$0.77200   &111&    -0.011$\pm$0.049     &    -0.066&     1.112&     2.539&     0.438&Accept~$H_{0}$&    -0.042\\
$J-K_{s}$&      1&  -0.14100$-$0.25700   &         100&    -0.257$\pm$0.057     &       ---&       ---&       ---&       ---&       ---         &       ---\\
& 2&   0.25700$-$0.33300   &100&     0.327$\pm$0.263     &    -0.257&     2.218&     2.544&     0.872&Accept~$H_{0}$&     0.252\\
& 3&   0.33300$-$0.41100   &100&    -0.208$\pm$0.238     &     0.252&     1.938&     2.544&     0.762&Accept~$H_{0}$&    -0.099\\
& 4&   0.41100$-$0.77200   &107&     0.100$\pm$0.052     &    -0.099&     3.787&     2.541&     1.491&Reject~$H_{0}$&     0.197\\
$H-K_{s}$&      1&  -0.13400$-$0.26100   &         100&    -0.143$\pm$0.049     &       ---&       ---&       ---&       ---&       ---         &       ---\\
& 2&   0.26100$-$0.33300   &100&     0.265$\pm$0.213     &    -0.143&     1.913&     2.544&     0.752&Accept~$H_{0}$&     0.164\\
& 3&   0.33300$-$0.40700   &100&    -0.071$\pm$0.187     &     0.164&     1.254&     2.544&     0.493&Accept~$H_{0}$&     0.048\\
& 4&   0.40700$-$0.77200   &115&     0.059$\pm$0.037     &     0.048&     0.313&     2.538&     0.123&Accept~$H_{0}$&     0.049\\
$V-J$&      1&  -0.14100$-$0.25600   &          99&     0.036$\pm$0.130     &       ---&       ---&       ---&       ---&       ---         &       ---\\
& 2&   0.25600$-$0.33300   &101&     0.187$\pm$0.654     &     0.036&     0.231&     2.543&     0.091&Accept~$H_{0}$&     0.050\\
& 3&   0.33300$-$0.41400   &100&     0.170$\pm$0.570     &     0.050&     0.211&     2.544&     0.083&Accept~$H_{0}$&     0.060\\
& 4&   0.41400$-$0.77200   &108&     0.426$\pm$0.137     &     0.060&     2.667&     2.540&     1.050&Reject~$H_{0}$&     0.444\\
$V-H$&      1&  -0.14100$-$0.25600   &          99&    -0.028$\pm$0.146     &       ---&       ---&       ---&       ---&       ---         &       ---\\
& 2&   0.25600$-$0.33300   &100&     0.204$\pm$0.757     &    -0.028&     0.307&     2.544&     0.121&Accept~$H_{0}$&    -0.000\\
& 3&   0.33300$-$0.41500   &101&    -0.262$\pm$0.581     &    -0.000&     0.450&     2.543&     0.177&Accept~$H_{0}$&    -0.046\\
& 4&   0.41500$-$0.77200   &104&     0.356$\pm$0.136     &    -0.046&     2.964&     2.542&     1.166&Reject~$H_{0}$&     0.423\\
$V-K_{s}$&      1&  -0.14100$-$0.25600   &         100&    -0.277$\pm$0.155     &       ---&       ---&       ---&       ---&       ---         &       ---\\
& 2&   0.25600$-$0.33300   &100&     0.525$\pm$0.778     &    -0.277&     1.031&     2.544&     0.405&Accept~$H_{0}$&     0.048\\
& 3&   0.33300$-$0.41400   &100&    -0.290$\pm$0.635     &     0.048&     0.533&     2.544&     0.209&Accept~$H_{0}$&    -0.023\\
& 4&   0.41400$-$0.77200   &105&     0.412$\pm$0.145     &    -0.023&     3.007&     2.541&     1.183&Reject~$H_{0}$&     0.491\\
$I-J$&      1&  -0.14100$-$0.26100   &         100&     0.022$\pm$0.083     &       ---&       ---&       ---&       ---&       ---         &       ---\\
& 2&   0.26100$-$0.33400   & 99&    -0.511$\pm$0.366     &     0.022&     1.457&     2.544&     0.573&Accept~$H_{0}$&    -0.284\\
& 3&   0.33400$-$0.42000   &101&     0.201$\pm$0.314     &    -0.284&     1.543&     2.543&     0.607&Accept~$H_{0}$&     0.010\\
& 4&   0.42000$-$0.77200   &103&     0.191$\pm$0.083     &     0.010&     2.178&     2.542&     0.857&Accept~$H_{0}$&     0.165\\
$I-H$&      1&  -0.14100$-$0.25800   &         100&    -0.079$\pm$0.092     &       ---&       ---&       ---&       ---&       ---         &       ---\\
& 2&   0.25800$-$0.33300   & 98&    -0.531$\pm$0.469     &    -0.079&     0.964&     2.545&     0.379&Accept~$H_{0}$&    -0.250\\
& 3&   0.33300$-$0.41600   &102&    -0.360$\pm$0.354     &    -0.250&     0.310&     2.543&     0.122&Accept~$H_{0}$&    -0.264\\
& 4&   0.41600$-$0.77200   &103&     0.131$\pm$0.081     &    -0.264&     4.860&     2.542&     1.912&Reject~$H_{0}$&     0.491\\
$I-K_{s}$&      1&  -0.14100$-$0.25600   &         100&    -0.308$\pm$0.104     &       ---&       ---&       ---&       ---&       ---         &       ---\\
& 2&   0.25600$-$0.33300   & 99&    -0.035$\pm$0.465     &    -0.308&     0.587&     2.544&     0.231&Accept~$H_{0}$&    -0.245\\
& 3&   0.33300$-$0.41500   &101&    -0.298$\pm$0.383     &    -0.245&     0.139&     2.543&     0.054&Accept~$H_{0}$&    -0.248\\
& 4&   0.41500$-$0.77200   &104&     0.203$\pm$0.092     &    -0.248&     4.903&     2.542&     1.929&Reject~$H_{0}$&     0.621\\
\hline
\end{tabular}
\end{center}
\end{minipage}
\end{table*}

\end{document}